\documentclass[reqno,a4paper]{amsart}
\usepackage{amssymb}
\usepackage{amsmath}
\usepackage{cite}

\makeatletter
\@addtoreset{equation}{section}
\makeatother

\renewcommand\epsilon{\varepsilon}

\newcommand{\id}{{1 \mskip -5mu {\rm I}}}
\newcommand{\mc}[1]{{\mathcal #1}}
\newcommand{\bb}[1]{{\mathbb #1}}

\begin{document}

\begin{titlepage}

\par\vskip 1cm\vskip 2em

\begin{center}

{\LARGE  {\bf
Large deviations for a stochastic \\~
\\
 model of heat flow }}

\par
\bigskip\bigskip\bigskip\bigskip

{\large
\begin{tabular}[t]{c}
$\mbox{Lorenzo Bertini}^{1}\,,
\quad\quad \mbox{Davide Gabrielli}^{2}\,,\quad\quad
\mbox{Joel L.\ Lebowitz}^{3}$
\\
\end{tabular}
\par
}
\bigskip\bigskip
{\small
\begin{tabular}[t]{ll}
{\bf 1} & Dipartimento di Matematica, Universit\`a di Roma La Sapienza\\
&  P.le A.\ Moro 2, 00185 Roma, Italy\\
{\bf 2} & Dipartimento di Matematica, Universit\`a dell'Aquila\\
&  67100 Coppito, L'Aquila, Italy \\
{\bf 3} & Department of Mathematics and Physics, Rutgers University \\
& New Brunswick, NJ 08903, USA\\
\end{tabular}
}
\end{center}
\bigskip
\noindent Dedicated to Mitchell Feigenbaum on the Occasion of His
Sixtieth Birthday

\bigskip
\vskip 1 em
\centerline{\bf Abstract}
\smallskip
\noindent We investigate a one dimensional chain of $2N$ harmonic
oscillators in which neighboring sites have their energies
redistributed randomly. The sites $-N$ and $N$ are in contact with
thermal reservoirs at different temperature $\tau_-$ and $\tau_+$.
Kipnis, Marchioro, and Presutti \cite{KMP} proved that this model
satisfies {}Fourier's law and that in the hydrodynamical scaling
limit, when $N \to \infty$, the stationary state has a linear energy
density profile $\bar \theta(u)$, $u \in [-1,1]$. We derive the
large deviation function $S(\theta(u))$ for the probability of
finding, in the stationary state, a profile $\theta(u)$ different
from $\bar \theta(u)$. The function $S(\theta)$ has striking
similarities to, but also large differences from, the
corresponding one of the symmetric exclusion process.  Like the
latter it is nonlocal and satisfies a variational equation. Unlike
the latter it is not convex and the Gaussian normal fluctuations
are enhanced rather than suppressed compared to the local
equilibrium state. We also briefly discuss more general model and find
the features common in these two and other models whose $S(\theta)$ is
known.

\vfill





\vskip 1.0em
\noindent

\end{titlepage}
\vfill\eject

\section{Introduction}

The properties of systems maintained in stationary nonequilibrium
states (SNS) by contacts with very large (formally infinite) thermal
reservoirs in different equilibrium states are of great theoretical
and practical importance.  These are arguably the simplest examples of
nonequilibrium systems to which the elegant, universal, and successful
formalism of equilibrium statistical mechanics might hopefully be
extended.  A striking universal feature of equilibrium systems is the
Boltzmann--Einstein relation according to which fluctuations in
macroscopic observables, arising from the grainy microscopic structure
of matter, can be described fully in terms of the macroscopic
thermodynamic functions (entropy, free energy) without any recourse to
the microscopic theory.
In trying to develop a similar formalism for SNS we have to start with
the fluctuations.  There has therefore been much effort devoted to
developing a mathematically rigorous fluctuation theory for simple
model SNS.  This has led to some interesting recent results for
conservative systems in contact with particle reservoirs at different
chemical potentials
\cite{BDGJL3,BDGJL1,BDGJL2,BG,DLS1,DLS2,DLS3}.

In particular it has been possible to obtain explicitly the large
deviation functionals (LDF) for some one dimensional lattice
systems. The internal dynamics of these systems is governed by
simple exclusion processes, symmetric (SEP) or asymmetric (ASEP),
while the entrance and exit of particles at the two boundaries are
prescribed by the chemical potentials, $\lambda_\pm$, of the right
and left reservoirs. The LDF gives the logarithm of the
probabilities of finding macroscopic density profiles $\rho(u)$,
where $u$ is the macroscopic space variable, different from the
typical values $\bar \rho(u)$; namely we have $\mathop{{\rm
Prob}}(\rho(u)) \sim \exp\{-N {\mathcal F}(\rho)\}$ where $N$ is
the number of lattice sites.

In the symmetric case, the situation we shall be primarily concerned
with here, the typical profile $\bar\rho(u)$ is given by the
stationary solution of the diffusion equation $\partial_t \rho(t,u) =
(1/2) \partial_u\big(D\partial_u \rho(t,u)\big)$, $u \in[-1,1]$ with
boundary conditions $\bar \rho(\pm 1) = \rho_\pm$.  The values
$\rho_\pm$ correspond to the densities in an equilibrium system with
chemical potentials $\lambda_\pm$.  The latter can be obtained by
setting the chemical potential of both end reservoirs equal to each
other, $\lambda_+ = \lambda_-$.  We note that in this equilibrium
case, the function ${\mathcal F}$ is simply related to the free energy
of the system.  {}For $\rho_+ \ne \rho_-$ and constant diffusion
coefficient $D$ (that is density independent and spatially uniform)
the profile $\bar \rho(u)$ is linear; this is the only case solved so
far for the SEP.  The results for the LDF of the SEP for this SNS
contained some surprises.

The most striking of these is non--locality: the probability of
density profiles $\rho_A(u)$ and $\rho_B(u)$ in disjoint
macroscopic regions $A$ and $B$ is not given by a product of the
separate probabilities, i.e.\ the LDF is not additive.  This is
very different from the equilibrium case where the LDF is given
(essentially) by an integral of the local free energy density for
the specified profiles $\rho_A(u)$ and $\rho_B(u)$, and is thus
automatically additive over macroscopic regions (even at critical
points).  Additivity is also true for the LDF of a system in full
local thermal equilibrium (LTE), e.g.\ for the stationary
nonequilibrium state of the zero range process. The microscopic
origin of the non--locality of the LDF for the open SEP lies in
the $O(N^{-1})$ corrections to LTE which extend over distances of
$O(N)$; $N$ is number of lattice sites, which goes to infinity in
the hydrodynamical scaling limit \cite{BJ,S}. So while the
deviations from LTE vanish in this limit their contributions to
the LDF, which involves summations over regions of size $N$, does
not

The effect of these $O(N^{-1})$ corrections to LTE is already present
at the level of Gaussian fluctuations about $\bar\rho(u)$.  These were
computed by Spohn in 1983 \cite{S} who found that the contributions
from the deviations from LTE made a finite contribution to the
variance of these Gaussian fluctuations, causing them to decrease, for
the SNS of the SEP from their LTE values.  The reduction in the
variance of Gaussian fluctuations can be recovered from the LDF
by setting $\rho(u) = \bar \rho(u) + N^{-1/2} \phi(u)$.
In fact in \cite{BDGJL1,DLS2} it is shown that ${\mathcal F}(\rho)$
for the SEP dominates the LDF coming from the corresponding LTE state
and therefore the fluctuations are suppressed.

The above observations about the SEP raise many questions about the
nature of the SNS of more realistic systems.  Do their LDF and
Gaussian fluctuations behave similarly to those of the SEP?  In
particular, to what extent do the LDF for SNS play a ``similar role''
to free energies in equilibrium systems?  In the absence of more
solved examples it is difficult to answer these questions. It is
therefore useful to find and investigate the SNS of other model
systems for which the LDF can be found and compare them to that of the
SEP.  This is what we do in the present paper and then discuss the
limited universality of the results.

The SNS we consider here is a simple
stochastic model of heat conduction in a crystal. It is well
known, see e.g.\ \cite{LS,RLL}, that harmonic chains do not obey
{}Fourier's law of heat conduction.  On the other hand, Kipnis,
Marchioro, and Presutti \cite{KMP} introduced a model of
mechanically uncoupled harmonic oscillators in which nearest
neighbor oscillators redistribute randomly their energy. This
system is then coupled to thermal reservoirs at different
temperatures and, thanks to the stochastic dynamics, the validity
of {}Fourier's law is proven. In particular the stationary energy
density $\bar\theta(u)$ is a linear profile as in the SEP. We
mention that a more sophisticated stochastic model of coupled
harmonic oscillators has been recently investigated. The evolution
is given by superimposing the Hamiltonian dynamics with a
stochastic one in which two nearest oscillators randomly exchange 
momenta. This model has two conservations laws (energy and total
length); the hydrodynamic limit is proven in \cite{B} for the
equilibrium case and in \cite{BO} for nonequilibrium, Gaussian
fluctuations are analyzed in \cite{FON}.

In this paper we consider the Kipnis--Marchioro--Presutti model, our
main result is the derivation of the corresponding LDF, that we denote
by $S(\theta)$.
It turns out that this function has both strong similarities and
significant differences from that of the SEP.  Like for the SEP the
LDF is nonlocal and yields Gaussian fluctuations about $\bar
\theta(u)$.  Unlike the SEP, however, it is obtained by minimization,
rather than maximization, of a ``proto LDF'' and the variance is
increased compared to that obtained from LTE.  Also in contrast to the
SEP the LDF, $S(\theta)$, is not convex.  We discuss these
similarities and differences in section \ref{s:7}, where we also give some
generalization of our and previous results to a larger class of model
systems.

\section{The model and main result}
\label{secmodel}

{}Following \cite{KMP} we consider a chain of one--dimensional harmonic
oscillators located at sites $x\in [-N,N]\cap\bb{Z}=:\Lambda_N$ and
described by the canonical coordinates $(q_x,p_x)$. The oscillators
are mechanically uncoupled so that the Hamiltonian of the chain is
$H=\sum_{x\in \Lambda_N}(p_x^2+q_x^2)/2$. The harmonic
oscillators are however coupled by the following stochastic dynamics.
Every pair of nearest neighbors sites waits an exponential time of rate
one and then the corresponding oscillators exchange energy.
More precisely, let $(q_y,p_y)$, $(q_{y+1},p_{y+1})$ be the canonical
coordinates at the sites $y$, $y+1$;
when the exponential clock between $y$ and $y+1$ rings
then the new values $(q'_y,p'_y)$, $(q'_{y+1},p'_{y+1})$ are distributed
according to the uniform distribution on the surface of constant
energy
$$
\frac{1}{2}\big[ (q'_y)^2+(p'_y)^2\big]
+\frac{1}{2}\big[ (q'_{y+1})^2+(p'_{y+1})^2\big] =
\frac{1}{2}\big[ q^2_y+p^2_y\big]
+\frac{1}{2}\big[ q^2_{y+1}+p^2_{y+1} \big]
$$
Moreover the boundary site $-N$, respectively $+N$,
waits an exponential time of rate one and then
the corresponding oscillator assume an energy distributed
according to a Gibbs distribution with temperature $\tau_-$,
respectively $\tau_+$.
All the exponential clocks involved in the dynamics are independent.

{}From a mathematical point of view it is sufficient to
look only at the local energy given by the
random variables $\xi_x:=\big( p_x^2+q_x^2\big)/2$, for which we get
a closed evolution described by the following Markov process.
The state space is $\Sigma_N := \bb{R}_+^{\Lambda_N}$, an element of
$\Sigma_N$ is denoted by $\xi:=\{\xi_x\,,\: x\in\Lambda_N\}$.
The infinitesimal generator of the process is the sum of a bulk
generator  $L_0$ plus two boundary generators $L_+$ and $L_-$
\begin{equation}
\label{gen}
L_N :=   N^2 \big[ L_0 + L_- + L_+ \big]
\end{equation}
in which we have speeded up the time by the factor $N^2$, this corresponds to
the diffusive scaling.

The bulk dynamics $L_0$ is defined as
$$
L_0 :=  \sum_{x=-N}^{N-1} L_{x,x+1}
$$
where
\begin{equation}
\label{gen1}
L_{x,x+1} f (\xi) :=
\int_0^1\!dp \:\big[ f( \xi^{(x,x+1),p} ) - f( \xi) \big]
\end{equation}
in which the configuration $\xi^{(x,x+1),p}$ is obtained from $\xi$ by
moving a fraction $p$ of the total energy across the bond $\{x,x+1\}$
to $x$ and a fraction $1-p$ to $x+1$, i.e.\
$$
(\xi^{(x,x+1),p})_y:=\left\{
\begin{array}{ll}
\xi_y & \textrm{ if } \ \ y\neq x,x+1 \\
p\, (\xi_x+\xi_{x+1}) & \textrm{ if  }\ \ y=x \\
(1-p)\, (\xi_x+\xi_{x+1}) & \textrm{ if }\ \ y=x+1 \\
\end{array}
\right.
$$

The boundary generators $L_\pm$ are defined by a heat bath dynamics
with respect to thermostats at temperatures $\tau_\pm$, i.e.\
$$
L_{\pm} f (\xi ) := \int_0^\infty\!\! dr \:
\frac{1}{\tau_{\pm}} e^{-r/\tau_{\pm}}
\big[ f ( \xi^{\pm N,r} ) - f ( \xi )  \big]
$$
in which the configuration $\xi^{\pm N,r}$ is obtained from $\xi$ by
setting the energy at $\pm N$ equal to $r$, i.e.\
$$
(\xi^{x,r})_y:=\left\{
\begin{array}{ll}
\xi_y  & \textrm{ if }\ \ y\neq x  \\
r & \textrm{ if  }\ \ y=x \\
\end{array}
\right.
$$
Note that we have set the Boltzmann constant equal to one. The process
generated by (\ref{gen}), denoted by $\xi(t)$, will be called the KMP
process.

We denote by $u\in[-1,1]$ the macroscopic space coordinates and
introduce the space of energy profiles as
$\mc M :=\{ \theta \in L_1([-1,1],du)\,:\: \theta(u)\ge 0 \}$.
We consider $\mc M$ equipped with the weak topology namely,
$\theta_n\to\theta$ iff for each continuous test function $\phi$ we
have $\langle \theta_n,\phi\rangle \to \langle \theta,\phi\rangle$,
where $\langle\cdot,\cdot \rangle$ is the inner product in $L_2([-1,1],du)$.
Given a microscopic configuration $\xi\in \Sigma_N$,
we introduce the empirical energy $\pi_N(\xi)$ by
mapping $\xi$ to the macroscopic profile
\begin{equation}
\label{eme}
\left[\pi_N (\xi )\right](u):=
\sum_{x=-N}^{N} \xi_x \:
\id_{\big[\frac{x}{N}- \frac{1}{2N}, \frac{x}{N}+ \frac{1}{2N} \big]} (u)
\end{equation}
note that $\pi_N(\xi)\in \mc M$ is a piecewise constant function.

In the case when $\tau_-=\tau_+=\tau$ it is easy to show that
$L_N$ is reversible with respect to the product of exponential
distribution with parameter $\tau$, i.e.\ the invariant measure is
given by the equilibrium Gibbs measure at temperature $\tau$,
\begin{equation}
d\mu_{N,\tau} (\xi ) = \prod_{x=-N}^{N} \frac{d\xi_x}{\tau} \:
e^{-\xi_x/\tau}
\end{equation}

When $\xi\in \Sigma_N$ is distributed according to $\mu_{N,\tau}$
then the empirical energy $\pi_N (\xi )$ concentrates, as $N\to\infty$
on the constant profile $\tau$ according to the following law of large
numbers. {}For each $\delta>0$ and each continuous test function
$\phi=\phi(u)$
\begin{equation}
\lim_{N\rightarrow \infty}
\mu_{N,\tau}\Big( \big|
\langle \pi_N(\xi ),\phi\rangle - \langle \tau,\phi\rangle
\big| > \delta \Big)=0
\label{llnr}
\end{equation}
where $\tau\in \mc M$ is the constant function with that value.

In this equilibrium case it is also easy to obtain the large deviation
principle associated at the law of large numbers (\ref{llnr}).
More precisely, the probability that the empirical energy $\pi_N(\xi)$
is close to some profile $\theta\in\mc M$ different from $\tau$ is
exponentially small in $N$ and given by a rate functional $S_0$
\begin{equation}
\label{ldpr}
\mu_{N,\tau}\left(\pi_N(\xi )\sim \theta\right)\asymp
\exp\big\{-N \, S_0(\theta) \big\}
\end{equation}
where $\pi_N(\xi )\sim \theta$ means closeness in the weak topology of
$\mc M$ and $\asymp$ denotes logarithmic equivalence as $N\rightarrow \infty$.
The functional $S_0$ is given by
\begin{equation}
S_0(\theta) = \int_{-1}^{1} \!du \:
\left[ \frac{\theta(u)}{\tau} - 1 - \log\frac{\theta(u)}{\tau} \right]
= \int_{-1}^1 \!du \: s_0(\theta(u), \bar \theta_0)
\label{equientr}
\end{equation}
where $\bar \theta_0 = \tau$ is the constant energy density profile
for $\tau_+=\tau_- = \tau$.
The above functional can in fact be obtained as the Legendre transform
of the pressure $G_0(h)$
$$
S_0(\theta)=\sup_{h}\big[ \langle \theta,h\rangle -G_0(h) \big]
$$
where $G_0$ is defined as
\begin{equation}
\label{press}
G_0(h) :=
\lim_{N\rightarrow \infty} \frac{1}{N}
\log \bb{E}_{\mu_{N,\tau}}
\Big( e^{N \langle h,\pi_{N}(\xi )\rangle} \Big)
= - \int_{-1}^1 \!du \:
\log [ 1- \tau \, h(u)]
\end{equation}
in which $\bb{E}_{\mu_{N,\tau}}$ denotes the expectation with respect to
$\mu_{N,\tau}$.

\medskip
If $\tau_-\neq \tau_+$ the process generated by $L_N$
is no longer reversible and its invariant measure $\mu_{N,\tau_\pm}$
is not explicitly known. Theorem 4.2 in \cite{KMP} implies however
the following law of large numbers. {}For each $\delta>0$ and each
continuous $\phi$
\begin{equation}
\label{lln}
\lim_{N\rightarrow \infty}
\mu_{N,\tau_\pm}\Big(
\big| \langle \pi_N(\xi ),\phi\rangle-
\langle \bar{\theta}, \phi\rangle \big|> \delta\Big)=0
\end{equation}
where $\bar{\theta}$ is the linear profile interpolating $\tau_-$ and
$\tau_+$, i.e.\
\begin{equation}
\label{bart}
\bar{\theta}(u)= \tau_-\, \frac{1-u}{2}  \, +\,  \tau_+ \, \frac{1+u}{2}
\end{equation}

It is natural to look for
the large deviations asymptotic for $\mu_{N,\tau_\pm}$.
In the case of the symmetric simple
exclusion process (SEP) this program has been carried out in
\cite{BDGJL1,BDGJL2,DLS1,DLS2}. The main result of this paper is an
expression for the large deviation rate functional for $\mu_{N,\tau_\pm}$
analogous to the one for the SEP. The functional we obtain is
nonlocal,
as is the one for the SEP, but it turns out to be nonconvex while the
one for SEP is convex. We mention that non convexity of the rate
functional also occurs for the asymmetric exclusion process
\cite{DLS3}.

Without loss of generality we assume $\tau_-<\tau_+$ and introduce the set
$\mc T_{\tau_\pm} :=\{ \tau\in C^1([-1,1])\,:\: \tau'(u) > 0\,,\:
\tau(\pm 1) = \tau_\pm \}$, here $\tau'$ is the derivative of $\tau$.
Given $\theta\in \mc M$ and $\tau\in \mc T_{\tau_\pm}$
we introduce the trial functional
\begin{equation}
\mathcal{G}(\theta,\tau) := \int_{-1}^{1} \!du \:
\Big[ \frac{\theta(u)}{\tau(u)} - 1 - \log\frac{\theta(u)}{\tau(u)}
-\log \frac{ \tau'(u)}{[\tau_+-\tau_-]/2} \Big]
\label{funcG}
\end{equation}
In this paper we show that the empirical energy for $\mu_{N,\tau_\pm}$
satisfies the large deviation principle
with a nonlocal, nonconvex rate functional $S(\theta)$ given by
\begin{equation}
S(\theta) =\inf_{\tau\in \mathcal{T}_{\tau_\pm}} \mathcal{G}(\theta,\tau)
\label{infG}
\end{equation}
that is we have
\begin{equation}
\label{ldp}
\mu_{N,\tau_\pm}\left(\pi_N(\xi )\sim \theta\right)\asymp
\exp\big\{-N \, S(\theta) \big\}
\end{equation}

We note there is a very close similarity between (\ref{infG}) and the
analogous result for the SEP, we emphasize however that in
(\ref{infG}) we minimize over the auxiliary profile $\tau$, while in
SEP one needs to maximize. This is, of course, related to the non
convexity of our $S$ versus the convexity of the rate functional for
SEP. It would be very interesting to understand this basic difference
also in terms of the combinatorial methods in \cite{DLS1,DLS2,DLS3} besides
the dynamical approach presented here.

Given $\theta\in\mc M$, we show that the minimizer in (\ref{infG})
is uniquely attained for some profile
$\tau(u)=\tau[\theta](u)$; therefore
$S(\theta )=\mathcal{G}(\theta,\tau[\theta ])$.
Moreover $\tau[\theta](u)$ is the unique strictly increasing solution
of the boundary value problem
\begin{equation}\label{Deq}
\left\{
  \begin{array}{l}
{\displaystyle \vphantom{\Bigg\{}
\tau^2 \frac{\tau''}{(\tau')^2} + \theta -\tau
  = 0 } \\
\tau(\pm 1) = \tau_\pm
  \end{array}
\right.
\end{equation}
which is the Euler--Lagrange equation $\delta \mathcal{G} /\delta \tau
=0$ when $\theta$ is kept fixed.

We note that for $\theta=\bar\theta$ the solution of (\ref{Deq}) is
given by $\tau[\bar\theta]=\bar\theta$ therefore
$S(\bar\theta)=\mc G(\bar\theta,\bar\theta)=0$.
On the other hand, by the convexity of the real functions $\bb R_+\ni
x\mapsto x-1-\log x$ and $\bb R_+\ni x\mapsto - \log x$,
for each $\theta\in\mc M$ and $\tau\in \mc T_{\tau_\pm}$ we have
$\mc G (\theta,\tau)\ge 0$ hence $S(\theta) \ge 0$. By the same
argument we also get that $S(\theta) = 0 $ if and only if
$\theta=\bar\theta$. This shows that the large deviation principle
(\ref{ldp}) implies the law of large numbers (\ref{lln}) and gives an
exponential estimate as $N\to\infty$.
We finally remark that the reversible case (\ref{equientr}) is
recovered from (\ref{funcG})--(\ref{ldp}) in the limit
$\tau_+-\tau_-\to 0$ which impose $\tau(u)$ constant.

\bigskip
\noindent\emph{Outline of the following sections.}

\smallskip
Our derivation of the rate functional $S$ follows the
dynamical/variational approach introduced in \cite{BDGJL1,BDGJL3}.
We look first, in Section \ref{s:db},
at the dynamical behavior in the diffusive scaling limit in a
bounded time interval $[0,T]$.
In particular, we obtain a dynamical large deviation principle
which gives the exponential asymptotic for the event in which the
empirical energy follows a prescribed space--time path.

In Section \ref{s:qp} we introduce the quasi potential, it is defined
by the minimal cost, as measured by the dynamical rate functional, to
produce an energy fluctuation $\theta$ starting from the typical
profile $\bar\theta$. By the arguments in \cite{BDGJL1,BDGJL3}, the
quasi potential equals the rate functional $S(\theta)$ of the
invariant measure $\mu_{N,\tau_\pm}$. A mathematical rigorous proof of
this statement for the SEP is given in \cite{BG}.
As discussed in \cite{BDGJL1,BDGJL3}, the quasi potential is the
appropriate solution of a Hamilton--Jacobi equation which involves the
transport coefficients of the macroscopic dynamics.
The derivation of the functional $S$ is then completed by showing that
(\ref{infG}) is the appropriate solution of this Hamilton--Jacobi
equation. As in the case of the SEP we are also able, by following this
dynamical/variational approach,  to characterize the
minimizer for the variational problem defining the quasi potential; this
path is the one followed by the process, with probability going to one as
$N\to\infty$, in the spontaneous creation of the fluctuation $\theta$.
In Section \ref{s:qp} we also show that the functional $S$ is not
convex, obtain its expression for constant profiles $\theta$, and
derive an \emph{additivity principle} analogous to the one for simple
exclusion processes obtained in \cite{DLS2,DLS3}.

In the remaining part of the paper we discuss some extensions of the
previous results. In particular, in Section \ref{s:d>1} we discuss the
KMP process in higher space dimension, $d\ge 1$, and obtain an upper
bound for the quasi potential in terms of the local equilibrium one.
We note that for the SEP it is possible to prove \cite{BDGJL1,BDGJL2} an
analogous {\em lower bound}. We also discuss the Gaussian fluctuations
around the stationary profile $\bar\theta$; as for the SEP
\cite{BDGJL1,DFIP,DLS1,DLS2,S}
the correction due to nonequilibrium
is given by the Green function of the Dirichlet Laplacian. In
particular, this correction is non local; as in the case of the SEP,
this is due to the long range correlations \cite{KMP}. However, for
the KMP process, the nonequilibrium enhances the Gaussian
fluctuations while in the SEP it decreases them. As the covariance of
the Gaussian fluctuations equals the inverse of the second
derivative of $S(\theta)$ at $\bar\theta$, the enhancement of Gaussian
fluctuations corresponds to the upper bound of $S(\theta)$ in terms of
the local equilibrium functional.
In the analysis in \cite{KMP} a crucial role is played by a process,
in \emph{duality} with respect to the KMP process, in which the local
variable at the site $x$ takes integral values. In Section \ref{s:du}
we discuss briefly the large deviations properties of this dual model
and obtain the expression for the large deviation functional.
{}Finally in Section \ref{s:7} we discuss the derivation of the large
deviation functional for generic one--dimensional nonequilibrium
symmetric models with a single conservation law. We obtain a simple
condition, which is satisfied by the zero range process, the
Ginzburg--Landau dynamics, the SEP, the KMP process and its
dual, that allows the derivation of the large deviation function by
means of a suitable trial functional.
Even when this condition fails to hold, it yields a simple criterion to
predict the enhancement/supression of the Gaussian fluctuation in the
stationary nonequilibrium state with respect to the full local
thermal equilibrium.

\medskip
The discussion in this paper will be kept at the physicists level
of mathematical rigor. However, for the more mathematically
inclined reader, we shall point out the main differences and
technical difficulties with respect to the case of the SEP, which
has been analyzed in full mathematical rigor \cite{BDGJL2}.

\section{Macroscopic dynamical behavior}
\label{s:db}

In this Section we consider the KMP process in a bounded time interval
$[0,T]$ under the diffusive scaling limit. We discuss the law of large
numbers (hydrodynamic limit) and the associated dynamical large
deviations principle for the empirical energy (\ref{eme}).

Given a continuous strictly positive
energy profile $\theta\in C([-1,1];\bb R_+)$,
we denote by $\nu_{\theta}^N$  the probability on
$\Sigma_N$ corresponding to a local equilibrium distribution (LTE)
with an energy
profile given by $\theta$. It is defined as
$$
d\nu_\theta^N (\xi) := \prod_{x=-N}^{N} d \nu_{\theta,x}^N(\xi_x)
$$
where
$$
d \nu^N_{\theta,x}
:= \frac{d\xi_x }{\theta(x/N)}
\exp\Big\{ - \frac{\xi_x}{\theta(x/N)} \Big\}
$$
Given two probability measures $\nu,\mu$ on $\Sigma_N$ we denote by
$h(\nu|\mu)$ the relative entropy of $\nu$ with respect to $\mu$, it
is defined as
$$
h(\nu|\mu) := \int\! d\mu(\xi) \: \frac{d\nu(\xi)}{d\mu(\xi)}
\log \frac{d\nu(\xi)}{d\mu(\xi)}
$$

We shall consider the KMP process with initial condition distributed
according to the product measure $\nu_{\theta_0}^N$ for some energy
profile $\theta_0$.  A straightforward computation then shows there
exists a constant $C$ (depending on $\theta_0$) such that for any $N$ we
have the relative entropy bound
\begin{equation}
\label{reb}
h(\nu_{\theta_0}^N | \nu_{\bar\theta}^N) \le C N
\end{equation}
where $\bar\theta$ is the stationary energy profile (\ref{bart}).
By the weak law of large numbers for independent variables we
also have that $\nu_{\theta_0}^N$ is associated to
the energy profile $\theta_0$ in the following sense.
{}For each $\delta>0$ and each continuous $\phi$
\begin{equation}
\label{t=0}
\lim_{N\rightarrow \infty}
\nu_{\theta_0}^N \Big(
\big|\langle\pi_N(\xi),\phi\rangle -
\langle\theta_0,\phi\rangle \big| > \delta \Big)= 0
\end{equation}

We remark that for the SEP it is possible (and convenient, see
\cite{BDGJL2}) to consider deterministic initial conditions.  {}For the
KMP process, as the ``single spin space'' $\bb R_+$ is not discrete,
such initial conditions do not satisfy the entropy bound (\ref{reb}),
which is required in the standard derivation, see e.g.\
\cite{KL,Slib}, of the hydrodynamic limit.  {}For this reason we have
chosen the initial condition distributed according to the product
measure $\nu_{\theta_0}^N$.  On the other hand, by the method
developed in \cite{L}, it should be also possible to consider
deterministic initial configurations.

We denote by $\bb{P}_{\nu_{\theta_0}^N}$ the distribution of the KMP
process when the initial condition is distributed according to
$\nu_{\theta_0}^N$.  The measure $\bb{P}_{\nu_{\theta_0}^N}$
is a probability on the space $D([0,T];\Sigma_N)$
of right continuous with left limit paths from $[0,T]$ to $\Sigma_N$.
The expectation with respect to $\bb{P}_{\nu_{\theta_0}^N}$ is denoted by
$\bb{E}_{\nu_{\theta_0}^N}$.

\subsection{Hydrodynamic limit}

Equation (\ref{t=0}) is the law of large number for the
empirical energy at time $t=0$; the hydrodynamic limit states that
for each macroscopic time $t\in [0,T]$ there exists an energy profile
$\theta(t)$ such we have the same law of large numbers
\begin{equation}
\label{hl}
\lim_{N\rightarrow \infty}\bb{P}_{\nu_{\theta_0}^N}
\Big( \big| \langle\pi_N(\xi(t)),\phi\rangle
- \langle \theta(t),\phi \rangle \big| > \delta\Big)=0
\end{equation}
{}Furthermore, we can obtain the energy profile $\theta(t)$ by solving
the hydrodynamic equation. {}For the KMP process (as for the SEP)
this is simply the the linear heat equation with boundary conditions
$\tau_\pm$, i.e.\ $\theta(t)=\theta(t,u)$ solves
\begin{equation}
\label{he}
\left\{
\begin{array}{lcl}
{\displaystyle
\partial_t \theta(t)}
&=& {\displaystyle \frac 12 \Delta \theta(t) }\\
{\displaystyle \vphantom{\bigg\{}
\theta(t, \pm1)}
&=&  {\displaystyle  \tau_\pm } \\
{\displaystyle \theta(0,u)}
&=&
{\displaystyle \theta_0(u)}
\end{array}
\right.
\end{equation}
where $\Delta$ is the Laplacian. Note that the stationary profile
$\bar\theta$ in (\ref{bart}) is the unique stationary solution of
(\ref{he}).

We give below a brief heuristic derivation, which is particularly
simple for the KMP process, of the hydrodynamic limit. We refer to
\cite{ELS1,ELS2} for a rigorous proof in the case of the so called gradient
nonequilibrium models with finite single spin state space; the
extension to the KPM process should not present additional problems.

\medskip
Let $\phi$ be a smooth function whose support is a subset of $(-1,1)$;
from the general theory of Markov processes, we have that
\begin{equation}
\frac{d}{dt}  \bb{E}_{\nu_{\theta_0}^N}
\big(  \langle\pi_N(\xi(t)),\phi\rangle\big)
= \bb{E}_{\nu_{\theta_0}^N}\big( L_N \langle\pi_N(\xi(t)),\phi \rangle\big)
\label{genmark}
\end{equation}
Since the support of $\phi$ is a strict subset of $[-1,1]$, only
$N^2L_0$ contributes to $L_N\langle\pi_N(\xi(t)),\phi\rangle$.
A simple computation shows that, when $y\neq \pm N$,
\begin{equation}
L_0 \xi_y = \frac 12 \big[ \xi_{y-1}+\xi_{y+1}-2\xi_y \big]
\label{Lh}
\end{equation}
we thus get
\begin{equation*}
\begin{array}{lcl}
{\displaystyle
\!\!
L_N\langle\pi_N(\xi(t)),\phi\rangle
}
& =  &
{\displaystyle
\frac{N^2}{2}
\sum_{x\in\Lambda_N}
\big[ \xi_{x-1}(t)+\xi_{x+1}(t)-2\xi_x(t)\big]
\int_{x/N - 1/(2N)}^{x/N + 1/(2N)}\!\!du \: \phi(u)
}
\\
& \approx &
{\displaystyle \vphantom{\Bigg\{}
\frac{1}{2N}\sum_{x\in\Lambda_N}\Delta_N \phi(x/N) \: \xi_x(t)
\approx \frac 12\langle \pi_N(\xi(t)),\Delta \phi\rangle
}
\end{array}
\end{equation*}
here $\Delta_N\phi(x/N):=N^2 \big[ \phi((x-1)/{N})+\phi((x+1)/{N})
-2\phi(x/{N})\big] $ is the discrete Laplacian.
The first step above comes from (\ref{Lh}) and
(\ref{eme}), the second step from discrete integration by parts and
last step from the regularity of $\phi$.

We have thus obtained the weak formulation of (\ref{he}); it remains
to show that also the boundary condition $\theta(t,\pm 1)= \tau_\pm$ is
satisfied. {}For this we need to use the boundary generators $N^2
L_\pm$. These are Glauber like dynamics accelerated by $N^2$ so that
the energy has well thermalized to its equilibrium value. We get
\begin{equation}
\bb{E}_{\nu_{\theta_0}^N}\left(\xi_{\pm N}(t)\right) \approx \tau_{\pm}
\label{termo}
\end{equation}

A standard martingale computation shows that, with a negligible error as
$N\to\infty$, $\pi_N(\xi(t))$ becomes non random. We can then remove
the expectation value in the previous equations and get (\ref{hl}).

\subsection{Dynamic large deviations}

We want next to obtain the large deviation principle associated to the
law of large number (\ref{hl}); more precisely we want to estimate the
probability that the empirical energy
$\pi_N(\xi(t))$ does not follow the solution of (\ref{he}) but remains
close to some prescribed path $\pi=\pi(t,u)$. This probability will be
exponentially small in $N$ and we look for the exponential rate. We
follow the classic procedure in large deviation theory: we perturb the
dynamics in such a way that the path $\pi$ becomes typical and compute
the cost of such a perturbation.

Let  $H=H(t,u)$ be a smooth function vanishing at the boundary, i.e.\
$H(t,\pm 1)=0$. We then consider the following time dependent
perturbations of the generators $L_{x,x+1}$ in (\ref{gen1})
\begin{equation}
L_{x,x+1}^H f (\xi) 
:= \int_0^1 \! dp  \:
e^{ [ H(t,x/N)-H(t,(x+1)/N)] [ p \xi_{x+1} - (1-p) \xi_x ]}
\big[ f( \xi^{(x,x+1),p}) - f( \xi) \big]
\nonumber
\end{equation}
Note that we have essentially just added a small drift $N^{-1}\nabla
H(t,x/N)$ in 
the energy exchange across the bond $\{x,x+1\}$.
We denote by $\bb{P}_{\nu_{\theta_0}^N}^H$ the distribution on the path space
$D([0,T];\Sigma_N)$ of this perturbed KMP process. As before
$\bb{E}_{\nu_{\theta_0}^N}^H$ is the expectation with respect to
$\bb{P}_{\nu_{\theta_0}^N}^H$.

The first step to obtain the dynamic large deviations is to derive the
hydrodynamic equation for the perturbed KMP process. We claim that for
each $t\in [0,T]$, each continuous $\phi$, and each $\delta>0$ we have
\begin{equation}
\lim_{N\rightarrow \infty}\bb{P}_{\nu_{\theta_0}^N}^H
\Big( \big|\langle\pi_N(\xi(t)),\phi\rangle
- \langle \theta(t),\phi \rangle\big| > \delta\Big)=0
\label{llnconH}
\end{equation}
where $\theta(t)=\theta(t,u)$ solves
\begin{equation}
\label{idroF}
\left\{
\begin{array}{lcl}
{\displaystyle
\partial_t \theta (t)
}
&=&
{\displaystyle
\frac 12 \Delta \theta (t) - \nabla \big( \theta(t)^2 \nabla H(t) \big)
}
\\
{\displaystyle \vphantom{\bigg\{} \theta(t, \pm 1)
}
&=&
{\displaystyle
 \tau_\pm
}
\\
{\displaystyle \theta(0,u)
}
&=&
{\displaystyle
\theta_0(u)
}
\end{array}
\right.
\end{equation}

The argument to justify (\ref{idroF}) is similar to the previous
one. Including the effect of the perturbation, the computation
following (\ref{Lh})  now becomes (as before $y\neq \pm N$)
\begin{equation*}
\begin{array}{l}
{\displaystyle
L_0^{H}\xi_y
}
 =
{\displaystyle
\int_0^1\! dp \:
e^{[ H(t,(y-1)/{N})-H(t,y/N)] [ p \xi_{y}-(1-p)\xi_{y-1} ]}
\big[(1-p)(\xi_{y}+\xi_{y-1})-\xi_y\big]
}
\\
\quad\quad\quad
{\displaystyle \vphantom{\Bigg\{}
+\: \int_0^1\! dp\:
e^{[ H(t,y/N)-H(t,(y+1)/N)] [ p \xi_{y+1}-(1-p) \xi_{y} ]}
\big[ p\,(\xi_{y+1}+\xi_{y})-\xi_y\big]
}
\\
\quad\quad
\approx \;
{\displaystyle \vphantom{\Bigg\{}
\frac{\xi_{y-1}+\xi_{y+1}-2\xi_y}{2}
+ \big[ H(t,(y-1)/{N}) - H(t,{y}/{N}) \big]
\frac {\xi_y\xi_{y-1}-\xi_y^2-\xi_{y-1}^2}{3}
}
\\
\quad\quad\quad
{\displaystyle
+\:  \big[ H(t,{y}/{N}) -H(t,(y+1)/{N}) \big]
\frac{ -\xi_y\xi_{y+1}+\xi_y^2+\xi_{y+1}^2 }{3}
}
\end{array}
\end{equation*}

As before, we consider a smooth function $\phi$ whose support is a
strict subset of $(-1,1)$; then only $N^2L_0^H$ contributes to
$L_N^H\langle\pi_N(\xi(t)),\phi\rangle$ and we get
\begin{equation*}
\begin{array}{l}
{\displaystyle
L_N^H \langle \pi_N(\xi(t)),\phi\rangle  \approx
\frac{1}{N}\sum_{x}
N^2 \phi(x/N) \bigg\{
\frac{ \xi_{x-1}(t)+\xi_{x+1}(t)-2\xi_x(t) }{2}
} \\
\quad\quad\quad
{\displaystyle \vphantom{\bigg\{}
+\:
\big[H(t,(x-1)/{N}) -H(t,{x}/{N})\big]
\frac{ \xi_x(t) \xi_{x-1}(t) -\xi_x(t)^2- \xi_{x-1}(t)^2}{3}
}
\\
\quad\quad\quad
{\displaystyle
+\: \big[ H(t,{x}/{N}) -H(t,(x+1)/{N})\big]
\frac{ -\xi_x(t) \xi_{x+1}(t) +\xi_x(t)^2+\xi_{x+1}(t)^2}{3}
\bigg\}
}
\\
\quad\quad
{\displaystyle
\approx\;
\frac{1}{N} \sum_{x}\xi_x(t) \Delta_N\phi(x/N)
}
\\
\quad\quad\quad
{\displaystyle
+\:
 \frac{1}{N}
\sum_{x}\frac{ -\xi_x(t)\xi_{x+1}(t)+\xi_x(t)^2+\xi_{x+1}(t)^2 }{3}
\:\nabla_N H(t,x/N) \: \nabla_N\phi(x/N)
}
\end{array}
\end{equation*}
where $\nabla_N f(x/N):=N[ f((x+1)/{N})-f({x}/{N})]$ is the
discrete gradient.  In the above computations we just used Taylor
expansions and discrete integrations by parts.
With respect to the very simple case discussed before, we face now
the main problem in establishing the hydrodynamic limit: the above
equation is not closed 
in $\pi_N(\xi(t))$, i.e.\ its right hand side is not a function of
$\pi_N(\xi(t))$.
In order to derive the hydrodynamic equation (\ref{idroF}), we need to
express $-\xi_x\xi_{x+1}+\xi_x^2+\xi_{x+1}^2$ in terms of the
empirical energy $\pi_N(\xi)$. This will be done by assuming a ``local
equilibrium'' state, we refer to \cite{BDGJL2,ELS1,ELS2,KL,Slib} for a
rigorous justification in the context of conservative interacting
particle systems.

Let us consider a microscopic site $x$ which is far from the boundary
and introduce a volume $V$, centered at $x$, which
is very large in microscopic units, but still infinitesimal at the
macroscopic level. The time evolution in $V$ is essentially given only
by the bulk dynamics $N^2 L_{0}^H$; since the total amount of energy in
$V$ changes only via boundary effects and we are looking at what
happens after $O(N^2)$ microscopic time units, we expect that the
system in $V$ has 
relaxed to the micro--canonical state corresponding  to the local
empirical energy $\pi_N(\xi(t))(x/N)$.
To compute this state let us construct first the canonical measure in $V$
with constant temperature $\tau>0$, namely the product measure
$d\nu_{V,\tau} (\xi ) := \prod_{x\in V} \tau^{-1} \, d\xi_x\: e^{-\xi_x/\tau}$.
Let now $m_{V,\theta}$ be the associated micro--canonical measure with
energy density $\theta$, i.e.\
$$
m_{V,\theta} (d\xi) := \nu_{V,\tau}
\Big( \, d\xi \: \Big| \sum_{x\in V} \xi_x =\theta |V| \Big)
$$
We introduce the function $\sigma(\theta)$ defined by
\begin{equation}\label{f-1}
\sigma(\theta) :=\lim_{V \uparrow \bb Z}\;
\bb{E}_{m_{V,\theta}} \big( -\xi_x\xi_{x+1}+\xi_x^2+\xi_{x+1}^2 \big)
\end{equation}
where we recall that $\bb{E}_{m_{V,\theta}}$ denotes the expectation with
respect to the probability $m_{V,\theta}$.
By the equivalence of ensemble we can compute $\sigma(\theta)$ also as
$$
\sigma(\theta )=\bb{E}_{\nu_{V,\theta}}
\big( -\xi_x\xi_{x+1}+\xi_x^2+\xi_{x+1}^2 \big)=3\, \theta^2
$$
According to the previous discussion, the system in the volume $V$ is
well approximated by a micro--canonical state with energy density
$\pi_N(\xi(t))(x/N)$.  As it is shown by the standard proofs in
hydrodynamic limits, see e.g.\ \cite{KL,Slib}, we can thus replace,
for $N$ large,
$-\xi_x(t)\xi_{x+1}(t)+\xi_x(t)^2+\xi_{x+1}(t)^2$ with
$3 [\pi_N(\xi(t))(x/N) ]^2$. We then obtain
\begin{equation}\label{drho}
\frac {d}{dt}  \bb{E}_{\nu_{\theta_0}^N} \big(
\langle \pi_N(\xi(t)), \phi\rangle   \big)
\approx \frac{1}{2} \langle \pi_N(\xi(t)),\Delta \phi\rangle
+ \langle \pi_N(\xi(t))^2 \nabla H, \nabla \phi\rangle
\end{equation}
which is the weak formulation of (\ref{idroF}). The arguments to show
that the boundary conditions $\theta(t,\pm 1)=\tau_\pm$ are satisfied and
to remove the expectation value are the same ones as in the derivation of
(\ref{he}).

\medskip

Let $\pi=\pi(t,u)$, $(t,u)\in[0,T]\times[-1,1]$ be a given path.
We recall that our task is to estimate the probability that the
empirical energy $\pi_N(\xi(t))$ is close to $\pi(t)$ (short for
$\pi(t,u)$). We write this probability in terms of the perturbed
KMP process, namely
\begin{equation}\label{perunper0}
\bb{P}_{\nu_{\theta_0}^N} \big( \pi_N(\xi(t))\sim \pi (t), t \in [0,T]\big)
=
\bb{E}_{\nu_{\theta_0}^{N}}^H
\bigg( \frac{d \bb P_{\nu_{\theta_0}^{N}}}{d \bb P_{\nu_{\theta_0}^{N}}^H}\:
\id_{\{\pi_N(\xi(t))\sim \pi(t) \}} \bigg)
\end{equation}
Equation (\ref{idroF}) tells us for which $H$ the
path $\pi$ becomes typical for the perturbed KMP process. We thus
choose $H(t,u)$ so that
\begin{equation}
\label{chooseH}
\left\{
\begin{array}{l}
{\displaystyle
\nabla \big( \pi(t)^2 \nabla H(t) \big)
= - \partial_t \pi (t) + \frac 12 \Delta \pi(t)
}
\\
{\displaystyle
H(t,\pm 1) = 0
}
\end{array}
\right.
\end{equation}
which is essentially a Poisson equation for $H$ (recall that $\pi$ is
fixed). With this choice we have, for $N$ large,
$\bb{P}_{\nu_{\theta_0}^N}^H \big( \pi_N(\xi(t))\sim \pi (t) \big) \approx 1$
and to derive the dynamical large deviation principle we only need to
compute the Radon--Nykodim derivative
$d \bb P_{\nu_{\theta_0}^{N}}/ d \bb P_{\nu_{\theta_0}^{N}}^H$.

We consider first the case of a deterministic initial configuration
$\xi_0\in\Sigma_N$. In this case, by a standard computation in the theory of
jump Markov processes, see e.g.\ \cite[Appendix 1.7]{KL} or
\cite[Appendix A]{BDGJL1}), we have
\begin{equation*}
\frac{d\bb{P}_{\xi_0}}{d\bb{P}_{\xi_0}^H}(\xi)  =
\exp\big\{ - N \mathcal{J}_{[0,T]}^N(\xi,H) \big\}
\end{equation*}
where
\begin{equation*}
\begin{array}{l}
{\displaystyle
\mathcal{J}_{[0,T]}^N(\xi,H)
:=
\langle \pi_N(\xi(T)),H(T)\rangle
-\langle \pi_N(\xi_0),H(0)\rangle
- \int_0^T\!dt \:
\langle\pi_N(\xi(t)),\partial_t H(t )\rangle
}
\\
\quad \quad \quad
{\displaystyle
- \: N^2\sum_{x=-N}^{N-1}\int_0^T\!dt
\int_0^1\! dp \: \Big\{
e^{[ H(t,{x}/{N})-H(t,(x+1)/{N})] \,
[ p \, \xi_{x+1}(t)-(1-p)\xi_x(t)] } - 1\Big\}
}
\end{array}
\end{equation*}
By Taylor expansion we then get
\begin{equation*}
\begin{array}{l}
{\displaystyle
\mathcal{J}_{[0,T]}^N(\xi,H )
\approx
\langle \pi_N(\xi(T)),H(T)\rangle
- \langle \pi_N(\xi_0),H(0)\rangle
- \int_0^T\!dt \:
\langle\pi_N(\xi(t)),\partial_t H(t )\rangle
}
\\
\quad\quad\quad\quad\quad
{\displaystyle
- \: \int_0^T \!dt \:
\frac{1}{2N}\sum_{x=-N+1}^{N-1}\xi_x(t)\Delta_N H(t,x/N)
}
\\
\quad\quad\quad\quad\quad
{\displaystyle \vphantom{\Bigg\{}
- \:
\frac{1}{2}\int_0^T\!dt \: \xi_{-N}(t)\,
N \big[  H(t,-1 + 1/N) - H(t,-1) \big]
}
 \\
\quad\quad\quad\quad\quad
{\displaystyle \vphantom{\Bigg\{}
+\:  \frac{1}{2}\int_0^T\!dt \:
\xi_{N}(t) \, N \big[ H(t,1)- H(t,1 -1/N)\big]
}
\\
\quad\quad\quad\quad\quad
{\displaystyle
-\:
\int_0^T\!dt \:
\frac{1}{2N}\sum_{x=-N}^{N-1}
\frac{ -\xi_x(t)\xi_{x+1}(t)+\xi_x^2(t)+\xi_{x+1}^2(t)}{3}
\: \big[ \nabla_N H(t,x/N)\big]^2
}
\end{array}
\end{equation*}
By the same argument given in the derivation of the perturbed hydrodynamic
equation  (\ref{idroF}), we can replace
$-\xi_x(t)\xi_{x+1}(t)+\xi_x(t)^2+\xi_{x+1}(t)^2$ by
$3 [ \pi_N(\xi (t))( x/N)]^2$.
Recalling that in (\ref{perunper0}) there is the indicator of the
event in which  $\pi_N(\xi(t))$ is close to $\pi(t)$, we get
\begin{equation*}
\begin{array}{lcl}
{\displaystyle
\mathcal{J}_{[0,T]}^N(\xi,H )
}
& \!\! \approx & \!\!
{\displaystyle
J_{[0,T]}(\pi)
\; = \;
\langle \pi(T),H(T) \rangle
-\langle \pi(0),H(0)\rangle
- \int_0^T\!dt\:
\langle\pi(t),\partial_t H(t)\rangle
}
\\
&& \!\!
{\displaystyle \vphantom{\Bigg\{}
 -\: \frac{1}{2} \int_0^T \!dt \: \langle \pi(t), \Delta H(t) \rangle
-\frac{1}{2}\int_0^T\!dt\:
\langle \pi(t)^2, [\nabla H(t)]^2 \rangle
}
\\
&& \!\!
{\displaystyle
-\: \frac{1}{2}\int_0^T\!dt \: \tau_- \nabla H(t,-1)
+ \frac{1}{2}\int_0^T\!dt \: \tau_+ \nabla H(t,1)
}
\end{array}
\end{equation*}
where we used the fact that the value of $\pi$ is fixed at the boundary,
$\pi(t,\pm 1)= \tau_\pm$.
Recalling that the perturbation $H$ has been chosen as the solution of
(\ref{chooseH}), integration by parts shows that
\begin{equation}
\label{finalaction}
J_{[0,T]}(\pi) = \frac 12 \int_0^T\!dt \:
\langle \nabla H (t), \pi(t)^2 \nabla H(t) \rangle
\end{equation}

To complete the derivation of the dynamical large deviation
functional, we only need to consider the fluctuations of the initial
condition. Recalling that we have chosen the initial condition
distributed according to the product measure $\nu_{\theta_0}^N$, a
straightforward computation on product measures (the one
carried out in (\ref{ldpr})--(\ref{press})) shows that
\begin{equation*}
\nu_{\theta_0}^N\big( \pi_N(\xi) \sim \pi(0) \big) \asymp
\exp\big\{ - N S_0 (\pi(0)| \theta_0)\big\}
\end{equation*}
where $S_0 (\pi(0)| \theta_0)$, which represents
the contribution to the dynamic large deviation from the initial
condition, is given by
\begin{equation}
\label{finalaction0}
S_0(\pi(0)| \theta_0) =
\int_{-1}^{1} \!du \:
\left[ \frac{\pi(0,u)}{\theta_0(u)} - 1 -
\log\frac{\pi(0,u)}{\theta_0(u)} \right]
\end{equation}

By collecting all the computations performed we finally get the
dynamical large deviation principle
\begin{equation}
\label{dldp}
\bb{P}_{\nu_{\theta_0}^N}
\Big( \pi_N(\xi (t) )\sim \pi (t), t\in [0,T] \Big)
\asymp \exp\big\{ -N \, I_{[0,T]}(\pi |\theta_0) \big\}
\end{equation}
where
\begin{equation}
\label{dldpbis}
I_{[0,T]}(\pi |\theta_0) = S_0 (\pi(0)| \theta_0) + J_{[0,T]} (\pi)
\end{equation}
We note again that $S_0 (\pi(0)| \theta_0)$ represents the cost to
create a fluctuation at time zero whereas $J_{[0,T]} (\pi)$ represents
the dynamical cost to follow the path $\pi(t)$ in the time interval
$[0,T]$. In the case of deterministic initial conditions, as the one
discussed in \cite{BDGJL2} for the SEP, we would have
$S_0(\pi(0)|\theta_0)=+\infty$ unless $\pi(0)=\theta_0$.

\medskip

\subsection {Remarks}

We conclude this section with some remarks on the rigorous derivation
of the dynamical large deviation principle (\ref{dldp}).  The
probability estimates needed are (not surprisingly) more subtle than
discussed here.
In fact, while in the proof of the hydrodynamic limit it is enough to
show that we can replace
$-\xi_x(t)\xi_{x+1}(t)+\xi_x^2(t)+\xi_{x+1}^2(t)$ by
$3[\pi_N(\xi(t))(x/N)]^2$ with an error vanishing as $N\to\infty$, in
the proof of the large deviations we need such an error to be of
$o(e^{-CN})$. This can be proven by the so called super exponential
estimate, see \cite{KL,KOV}, which is the key point in the rigorous
approach. This estimate has been extended to the non equilibrium SEP
in \cite{BDGJL2}. {}For the KMP process there is the additional
complication of a unbounded single spin space.
In \cite{DV} the dynamical large deviation principle is proven for
the Ginzburg--Landau model; however for the KMP process the situation
is more troublesome because the mobility $\pi^2$ is unbounded and the
reference measure is only exponentially decaying for large $\pi$.
There is also another technical point which requires some care. In
the usual proofs of large deviations from hydrodynamic behavior,
one first obtains the lower bound for a neighborhood of strictly
positive smooth paths $\pi$ and then uses approximations arguments to
extend the lower bound to any open set.  The approximations arguments
used for the SEP, see \cite{KL,KOV} for the equilibrium case and
\cite{BDGJL2} for nonequilibrium, take full advantage of the fact
(special for the SEP) that $J_{[0,T]}(\pi)$ is a convex functional. In
order to prove the dynamic large deviation principle for the KMP
process a more robust approximation method, possibly analogous to the
one in \cite{QRV}, is required.

\section{The quasi potential and its properties}
\label{s:qp}
In this Section we introduce the quasi potential, which measures the
minimal cost to produce a fluctuation of the energy profile in the
stationary state, and shows that
it can be obtained by solving the one--dimensional non linear boundary
value problem (\ref{Deq}). We also characterize the most probable
path followed by the KMP process in the spontaneous creation of such a
fluctuation. We finally show that the functional $S$ is not convex and
derive an \emph{additivity principle} analogous to the one in
\cite{DLS2,DLS3}.

Given $T>0$ and a strictly positive smooth $\theta\in \mc M$, we
introduce the set of energy paths which connect $\bar\theta$ to
$\theta$ in a time interval $[-T,0]$, i.e.\ we define
\begin{equation}
\mathcal{E}_{\theta, T}:=\big\{ \pi=\pi (t,u) \,:\: \pi(-T,u)
=\bar\theta(u)\,,\: \pi(0,u) = \theta(u) \big\}
\end{equation}
where we recall that the stationary energy profile $\bar\theta$ has
been defined in (\ref{bart}). Paths $\pi\in\mc E_{\theta,T}$ must
also satisfy the boundary condition $\pi(t,\pm 1)=\tau_\pm$; in fact
it can be shown \cite{BDGJL2} that $J_{[-T,0]}(\pi) =+\infty$ if the
path $\pi$ does not satisfy this boundary condition.
The \emph{quasi potential} is then defined as
\begin{equation}
\label{quasipot}
V(\theta) := \inf_{T>0} \; \inf_{\pi\in \mc E_{\theta,T}} \;
J_{[-T,0]}(\pi)
\end{equation}
where we recall that the functional $J$ is given in
(\ref{finalaction}).
By the general arguments in \cite{BDGJL1}, see also the rigorous proof in
\cite{BG} for the SEP, we have that the rate functional $S(\theta)$
for the invariant measure $\mu_{N,\tau_{\pm}}$, see (\ref{ldp}),
coincides with the quasi potential, i.e.\ $S(\theta)=V(\theta)$.

\subsection{Solution of the Hamilton--Jacobi equation}
Recalling that the perturbation $H$ in (\ref{finalaction}) solves
(\ref{chooseH}), the variational problem (\ref{quasipot}) consists in
minimizing the action  that correspond to the Lagrangian
\begin{equation}
\label{lagrange}
\mathcal{L}(\theta,\partial_t \theta) =  \frac 12
\big\langle \nabla^{-1} (\partial_t \theta -\frac 12 \Delta \theta),
\frac{1}{\theta^2} \nabla^{-1} (\partial_t \theta -\frac 12 \Delta
\theta) \big\rangle
\end{equation}
The associated Hamiltonian is
\begin{equation}
\mathcal{H}(\theta, H) :=
\sup_{\zeta}\big\{ \langle H,\zeta\rangle -\mathcal{L}(\theta,\zeta)
\big\}
=
\frac{1}{2} \big\langle \nabla H, \theta^2 \nabla H \big\rangle
+ \frac{1}{2}  \big\langle  H, \Delta \theta\big\rangle
\end{equation}

Noting that $V$ is normalized so that $V(\bar\theta)=0$, we obtain, by a
classical result in analytic mechanics, that $V(\theta)$ solves
the Hamilton--Jacobi equation $\mathcal{H}(\theta, \frac{\delta V}{\delta
\theta})=0$, i.e.\
\begin{equation}\label{hj}
\Big\langle \nabla \frac{\delta V}{\delta \theta}, \theta^2
\nabla \frac{\delta V}{\delta \theta} \Big\rangle
+\Big\langle  \frac{\delta V}{\delta \theta}, \Delta \theta
\Big\rangle =0
\end{equation}
where ${\delta V}/{\delta \theta}$ vanishes at the boundary and
$\theta(\pm 1) = \tau_\pm$.
We look for a solution of (\ref{hj}) in the form
\begin{equation}\label{cvhj}
\frac{\delta V}{\delta \theta} = \frac {1}{\tau} - \frac {1}{\theta}
\end{equation}
for some function $\tau=\tau[\theta](u)$ to be determined satisfying
the boundary conditions $\tau(\pm 1) = \tau_\pm$.
By plugging (\ref{cvhj}) into (\ref{hj}) and elementary computations,
analogous to the ones for the SEP discussed in \cite{BDGJL1}, we get
\begin{equation}
\label{hjspec}
\begin{array}{lcl}
0 & = &
{\displaystyle
\Big\langle \nabla \Big(\frac {1}{\tau} - \frac{1}{\theta}\Big),
\theta^2
\nabla  \frac{1}{\tau} \Big\rangle
}
\\
& = &
{\displaystyle \vphantom{\Bigg\{}
-\Big\langle\nabla (\theta-\tau), \frac{\nabla \tau}{\tau^2} \Big\rangle
+\Big\langle \frac{ (\nabla \tau)^2}{\tau^4},
 (\theta^2-\tau^2) \Big\rangle
}
\\
& = &
{\displaystyle
\Big\langle \frac{\theta - \tau}{\tau^4},
\, \tau^2\Delta \tau + (\theta-\tau) (\nabla \tau)^2
\Big\rangle
}
\end{array}
\end{equation}
Therefore a solution of (\ref{hjspec}) is obtained when $\tau$
satisfies the non linear boundary value problem (\ref{Deq}). Let
us denote by $\tau[\theta]$ the solution of (\ref{Deq}); recall
the definition (\ref{funcG}) of the functional $\mc G
(\theta,\tau)$ and that, since (\ref{Deq}) is the associated
Euler--Lagrange equation for fixed $\theta$, we have $\left[\delta
\mc G/\delta \tau\right] (\theta,\tau[\theta])=0$. By a direct
computation we then get
\begin{equation}
\frac{\delta}{\delta\theta} \mc G(\theta,\tau[\theta]) =
\frac{\delta \mc G}{\delta\theta} (\theta,\tau[\theta])
+ \frac{\delta \mc G}{\delta\tau} (\theta,\tau[\theta])
\, \frac{\delta \tau[\theta]}{\delta\theta}
=
\frac{1}{\tau[\theta]} - \frac{1}{\theta}
\end{equation}
which shows that $V(\theta)=\mc G(\theta,\tau[\theta])$ is a solution
of the Hamilton--Jacobi equation (\ref{hj}). To complete the
derivation of (\ref{ldp}) we next show that $V(\theta)$ meets the
criterion in \cite[\S 2.6]{BDGJL1}, i.e.\ it is the ``right solution''
of the Hamilton--Jacobi equation, and that the infimum in (\ref{infG})
is uniquely attained for $\tau=\tau[\theta]$, the solution of
(\ref{Deq}).

\subsection{The exit path}
\label{s:ep}

The characterization of the optimal path for the variational problem
(\ref{quasipot}) can be carried out according to the general scheme in
\cite{BDGJL2}. Let $V(\theta) = \mc G (\theta,\tau[\theta])$ and
$\pi(t)$, $t\in[-T,0]$ a strictly positive smooth path such that
$\pi(0)=\theta$.  By using that $V(\theta)$ solves Hamilton--Jacobi
equation (\ref{hj}), a simple computation shows that
\begin{equation*}
\begin{array}{l}
{\displaystyle
J_{[-T,0]}(\pi)
}
=
{\displaystyle
\frac 12 \int_{-T}^0\! dt \:
\bigg\langle
\nabla^{-1} \Big[ \partial_t \pi - \frac 12 \Delta \pi
+ \nabla \Big( \pi^2 \nabla \frac{\delta V}{\delta \theta} (\pi)\Big)
- \nabla \Big( \pi^2 \nabla \frac{\delta V}{\delta \theta} (\pi)\Big)
\Big]\, ,
}
\\
\quad\quad\quad\quad\quad
{\displaystyle
\phantom{ \frac 12 \int_{-T}^0\! dt}
\frac{1}{\pi^2}
\nabla^{-1} \Big[ \partial_t \pi - \frac 12 \Delta \pi
+ \nabla \Big( \pi^2 \nabla \frac{\delta V}{\delta \theta} (\pi)\Big)
- \nabla \Big( \pi^2 \nabla \frac{\delta V}{\delta \theta} (\pi)\Big)
\Big]
\bigg\rangle
}
\\
\quad\quad\quad
=
{\displaystyle
\:
V(\theta) - V (\pi(-T))
+\frac 12 \int_{-T}^0\! dt \:
\bigg\langle
\nabla^{-1} \Big[ \partial_t \pi - \frac 12 \Delta \pi
+ \nabla \Big( \pi^2 \nabla \frac{\delta V}{\delta \theta} (\pi)\Big)
\Big]\, ,
}
\\
\quad\quad\quad\quad\quad
{\displaystyle
\phantom{V(\theta) - V (\pi(0)) \frac 12 \int_{-T}^0\! dt }
\frac{1}{\pi^2}
\nabla^{-1} \Big[
\partial_t \pi - \frac 12 \Delta \pi
+ \nabla \Big( \pi^2 \nabla \frac{\delta V}{\delta \theta} (\pi)\Big)
\Big]
\bigg\rangle
}
\end{array}
\end{equation*}
Since the last term above is positive, the optimal path $\pi^*$ for
the variational problem (\ref{quasipot}) solves
\begin{equation}
\label{pi*}
\partial_t \pi^* = \frac 12 \Delta \pi^*
- \nabla \Big( (\pi^*)^2 \nabla \frac{\delta V}{\delta \theta} (\pi^*)\Big)
=  - \frac 12 \Delta \pi^*
+\nabla \Big( \frac{(\pi^*)^2}{(\tau[\pi^*])^2} \nabla \tau[\pi^*] \Big)
\end{equation}
where $\tau[\pi^*]= \tau[\pi^*] (t,u)$ denotes the solution of
(\ref{Deq}) with $\theta$ replaced by $\pi^*(t)$.

Let us denote by $\theta^*(t)=\pi^*(-t)$, $t\in [0,T]$, the time
reversed of the optimal path $\pi^*$. It is then not difficult to
show, see \cite[Appendix B]{BDGJL1} for the analogous computation in
the case of the SEP, that $\theta^*(t)$ can be constructed by the
following procedure. Given $\theta=\pi^*(0)=\theta^*(0)$, first let
$\tau_0 = \tau[\theta]$ be the solution of (\ref{Deq}), then solve the
heat equation with initial condition $\tau_0$, i.e.\ let $\tau(t)$ be
the solution of
\begin{equation*}
\left\{
\begin{array}{lcl}
{\displaystyle
\partial_t \tau(t)}
&=& {\displaystyle \frac 12 \Delta \tau(t) }\\
{\displaystyle \vphantom{\bigg\{}
\tau(t, \pm 1)}
&=&  {\displaystyle  \tau_\pm } \\
{\displaystyle \tau(0,u)}
&=&
{\displaystyle \tau_0(u)}
\end{array}
\right.
\end{equation*}
and finally set
\begin{equation*}
\theta^*(t) = \tau (t) - \tau(t)^2
\frac{\Delta \tau(t)}{ [\nabla \tau (t)]^2 }
\end{equation*}

Since $\tau(t) \to \bar \theta$ as $t\to\infty$ we get $\pi^*(-T)\to
\bar\theta$ as $T\to \infty$, hence $V(\pi^*(-T))\to V(\bar\theta)=0$.
The identification of the solution of the Hamilton--Jacobi equation
with the quasi potential follows from the characterization of the
minimizer $\pi^*$ obtained before.  In particular $V(\theta)$
satisfies the criterion discussed in \cite[\S 2.6]{BDGJL1}.

\subsection{Solution of equation (\ref{Deq})}

The existence of a solution for the nonlinear boundary value problem
(\ref{Deq}) can be proven by the same strategy as in
\cite{BDGJL2,DLS2}. We write (\ref{Deq}) in the integral--differential
form
\begin{equation*}
\tau(u) = \tau_- + (\tau_+-\tau_-)\:
\frac{{\displaystyle
\int_{-1}^u\! dv \: \exp\Big\{ \int_{-1}^{v}\! dw \:
    \frac{[\tau(w) -\theta(w)] \tau'(w) }{\tau(w)^2} \Big\}}}
{{\displaystyle
\int_{-1}^1\!dv \: \exp\Big\{
\int_{-1}^{v}\!dw
\frac{[\tau(w)-\theta(w)]\tau'(w)}{\tau(w)^2 }\Big\}}}
\end{equation*}
Then a solution of (\ref{Deq}) is a fixed point of
the integral--differential operator $\mathcal{K}_{\theta}[\tau]$ defined as
\begin{equation*}
\mathcal{K}_{\theta}[\tau]\, (u) := \tau_- + (\tau_+-\tau_-)\:
\frac{{\displaystyle
\int_{-1}^u\! dv \: \exp\Big\{ \int_{-1}^{v}\! dw \:
    \frac{[\tau(w) -\theta(w)] \tau'(w) }{\tau(w)^2} \Big\}}}
{{\displaystyle
\int_{-1}^1\!dv \: \exp\Big\{
\int_{-1}^{v}\!dw
\frac{[\tau(w)-\theta(w)]\tau'(w)}{\tau(w)^2 }\Big\}}}
\end{equation*}

We consider the case in which $\theta$ is bounded, namely we assume that
$\|\theta\|:=\sup_u |\theta(u)|<+\infty$.
Recalling that $\tau$ must be strictly increasing and such that
$\tau(\pm 1)=\tau_{\pm}$, we get
\begin{equation*}
 -\frac{\|\theta \|\tau'}{\tau_-\tau}\leq
\frac{(\tau-\theta)\tau'}{\tau^2}\leq \frac{\tau' }{\tau}
\end{equation*}
which yields
\begin{equation*}
\frac{\tau_+-\tau_-}{2}
\Big( \frac{\tau_-}{\tau_+}\Big)^{1+\frac{\|\theta\|}{\tau_-}}
\leq \frac{d}{du} \mathcal{K}_{\theta}[\tau](u)
\leq \frac{\tau_+-\tau_-}{2}
\Big(\frac{\tau_+}{\tau_-}\Big)^{1+\frac{\|\theta\|}{\tau_-}}
\end{equation*}
It is now easy to show, see \cite{BDGJL2} for more details, that for
each $\theta\in\mc M$, the operator $\mathcal{K}_{\theta}[\tau] $ maps
a compact convex subset of $\mc T_{\tau_\pm}$ into itself.  Hence, by
Schauder's fixed point theorem, we conclude the proof of the existence
of solution to (\ref{Deq}).

Uniqueness of solution to (\ref{Deq}) can also be proved with a
slight variation of the argument in \cite{DLS2}.  Let us consider
two different increasing solutions of (\ref{Deq}) $\tau_1(u)$ and
$\tau_2(u)$.  If $\tau_1'(-1)=\tau_2'(-1)$ then uniqueness of the
Cauchy problem implies $\tau_1=\tau_2$.  On the other hand, if
$\tau_1'(-1)>\tau_2'(-1)>0 $ then we denote by $\bar{u}$ the leftmost
point in $(-1,1]$ such that $ \tau_1(\bar{u})= \tau_2(\bar{u})$.  The
point $\bar{u}$ exists because $ \tau_1(1)= \tau_2(1)$ moreover we
have that $\tau_1'(\bar{u})\leq \tau_2'(\bar{u})$. {}From (\ref{Deq}) we
get
$$
\frac{d}{du} \Big( \frac{\tau}{\tau'}\Big)=\frac{\theta}{\tau}
$$
which integrated gives
$$
\frac{\tau(u)}{\tau'(u)}=\frac{\tau_-}{\tau'(-1)}+
\int_{-1}^{u}\! dv \: \frac{\theta(v)}{\tau(v)}
$$
we then deduce
$$
\frac{\tau_1(\bar{u})}{\tau_1'(\bar{u})}
-\frac{\tau_2(\bar{u})}{\tau_2'(\bar{u})}
=\frac{\tau_-}{\tau_1'(-1)}-\frac{\tau_-}{\tau_2'(-1)}
+\int_{-1}^{\bar u}\! dv \: \theta(v)
\Big[ \frac{1}{\tau_1(v)} - \frac{1}{\tau_2(v)} \Big]
$$
Since $\tau_1'(-1)>\tau_2'(-1)$ and $\tau_1(v)\ge \tau_2(v)$ for
$v\in [-1,\bar u ]$,  the right hand side above is strictly negative.
Recalling that $\tau_1(\bar{u})=\tau_2(\bar{u})$ we get
$\tau_1'(\bar{u})> \tau_2'(\bar{u})$, the desired contradiction.

In order to prove that the infimum in (\ref{infG}) is uniquely attained
for $\tau=\tau[\theta]$, the solution of (\ref{Deq}),
we perform the change of variable $\tau=e^{\varphi}$.
We then get the functional
\begin{equation}
\label{Gtilde}
\widetilde{\mathcal{G}}(\theta ,\varphi )  :=  \mathcal{G}(\theta ,
e^{\varphi })
 =  \int_{-1}^1\! du \: \Big[ \theta(u) \, e^{-\varphi(u)} - 1 -
\log \theta(u) - \log\frac{\varphi'(u)}{[\tau_+- \tau_-]/2 }\Big]
\end{equation}
which is strictly convex in $\varphi$; this trivially implies the claim.

\subsection{Non convexity of the quasi potential}

As we mentioned before, in the case of the SEP the quasi potential can
be obtained by a variational problem analogous to (\ref{funcG}) where
one maximizes over the auxiliary profile
\cite{BDGJL1,BDGJL2,DLS1,DLS2}.
In such a case, since the functional $\mc G (\theta,\tau)$ is convex
in $\theta$ for fixed $\tau$, the rate functional $S(\theta)$ is
trivially convex in $\theta$.  In the case of the KMP process we need
instead to minimize over the auxiliary profile $\tau$, therefore there
is no reason to expect $S(\theta)$ to be convex.  We now show, by an explicit
computation, that the rate functional is
indeed not convex.

We mention that non convexity of the rate functional $S$ has been
shown for the asymmetric exclusion process \cite{DLS3}; in that
case however the functional is degenerate, in the sense that there are
infinitely many profiles for which $S$ vanishes. Therefore the
mechanism of the non convexity is somehow different from the one in
the KMP process, where $S(\theta)$ vanishes only at $\bar\theta$.

\smallskip
To prove the non convexity of the rate functional $S$ we shall exhibit
profiles $\theta$ and $g$ so that, by choosing $\epsilon$ small enough,
we have
\begin{equation}
\label{nc1}
S(\theta) =
S\Big(\frac 12 [\theta+\epsilon g] +\frac 12 [\theta-\epsilon g] \Big) >
\frac 12 S(\theta +\epsilon g) +\frac 12 S(\theta  -\epsilon g)
\end{equation}
Let $\tau=\tau[\theta]$ be the strictly increasing solution of the
boundary value problem (\ref{Deq}), then by using (\ref{infG}) for
$\epsilon$ small enough and any profile $f$ vanishing at the
boundaries, $f(\pm 1)=0$, we have
\begin{equation*}
\begin{array}{lcl}
{\displaystyle
S(\theta+ \epsilon g)
}
&\le &
{\displaystyle
\mc G (\theta+\epsilon g, \tau+ \epsilon f)
}
\\
&=&
{\displaystyle \vphantom{\Bigg\{}
\mc G (\theta, \tau)
+\epsilon \int_{-1}^1\!du \:
\Big\{ \Big( \frac{\theta}{\tau} -1\Big)
\Big( \frac{g}{\theta} -\frac{f}{\tau}  \Big)
-\frac{f'}{\tau'}
\Big\}
}
\\
&&
+\;
{\displaystyle
\frac{\epsilon^2}{2}
\int_{-1}^1\!du \:
\Big\{ \Big( 2 \frac{\theta}{\tau} -1\Big)
\frac{f^2}{\tau^2}
+\frac{g^2}{\theta^2}
- 2 \frac{g f}{\tau^2}
+\frac{(f')^2}{(\tau')^2}
\Big\}
+o(\epsilon^2)
}
\end{array}
\end{equation*}
where we brutally Taylor expanded (\ref{funcG}).

Since $S(\theta)=\mc G (\theta, \tau)$, the inequality (\ref{nc1})
will follow if we show that, for an appropriate choice of $f$ (recall
that $\tau=\tau[\theta]$ is the solution of (\ref{Deq})) we can make
the quadratic term in the previous equation strictly negative, i.e.
\begin{equation}
\label{nc3}
\int_{-1}^1\!du \:
\Big\{ \Big( 2 \frac{\theta}{\tau} -1\Big)
\frac{f^2}{\tau^2}
+\frac{g^2}{\theta^2}
- 2 \frac{g f}{\tau^2}
+\frac{(f')^2}{(\tau')^2}
\Big\} < 0
\end{equation}

Let us introduce the function
\begin{equation*}
h(u)
:= \left\{
\begin{array}{lcl}
{\displaystyle \vphantom{\Big\{_{B_B}}
\tau_- + \frac{64}{81} (\tau_+-\tau_-) (1-u^6)
}
&\textrm{ if } &
{\displaystyle
-1\le u \le -1/2
}
\\
{\displaystyle \vphantom{\Big\{^{B^B}}
\tau_+ + \frac{4}{27} (\tau_+-\tau_-) (u-1)
}
&\textrm{ if } &
{\displaystyle
-1/2<  u \le 1
}
\end{array}
\right.
\end{equation*}
Note that $h\in C^1([-1,1])$, $h(\pm 1)=\tau_\pm$ and $h$ is strictly
increasing.
We choose the profile $\theta$ as
\begin{equation}
\label{nc4}
\theta (u)= h(u) \Big[ 1- h(u) \frac{ h''(u)}{ h'(u)^2} \Big]
\end{equation}
Note that $\theta>0$, i.e.\ $\theta$ is an allowed profile in $\mc M$;
the corresponding solution of the boundary value problem (\ref{Deq})
is $\tau[\theta]=h$. We further choose $f(u)=(1-u^2) h'(u)$ (note that
$f$ vanishes at the boundaries as required) and $g= f
\theta^2/h^2$. With the above choices the left hand side of
(\ref{nc3}) equals
\begin{equation*}
\begin{array}{l}
{\displaystyle
\int_{-1}^1\!du \:
\Big\{ \Big( 2 \frac{\theta}{h} -1\Big)
\frac{f^2}{h^2}
-\frac{f^2\theta^2}{h^4}
+\frac{(f')^2}{(h')^2}
\Big\}
}
\\
\quad\quad
=\;
{\displaystyle \vphantom{\Bigg\{}
\int_{-1}^1\!du \:
\Big\{ -\Big( \frac{\theta}{h} -1\Big)^2
\frac{f^2}{h^2}
+\frac{(f')^2}{(h')^2}
\Big\}
}
\\
\quad\quad
=\;
{\displaystyle \vphantom{\Bigg\{}
\int_{-1}^1\!du \:
\Big\{ -\frac{(h'')^2}{(h')^4} f^2
+\frac{(f')^2}{(h')^2}
\Big\}
}
\\
\quad\quad
=\;
{\displaystyle \vphantom{\Bigg\{}
\int_{-1}^1\!du \:
\Big\{ -\frac{(h'')^2 (1-u^2)^2}{(h')^2}
+\frac{1}{(h')^2} \big[ -2u h' + (1-u^2) h'' \big]^2
\Big\}
}
\\
\quad\quad
=\;
{\displaystyle \vphantom{\Bigg\{}
4\int_{-1}^1\!du \:
\Big\{ u^2 - u(1-u^2) \frac{h''}{h'}
\Big\}
}
\\
\quad\quad
=\;
{\displaystyle
4\Big\{ \int_{-1}^1\!du \: u^2 - 5
 \int_{-1}^{-1/2}\!\! du \: (1-u^2)
\Big\}
\;=\; 4 \Big\{ \frac{2}{3} - \frac{25}{24} \Big\} <0
}
\end{array}
\end{equation*}
This completes the proof of (\ref{nc3}) and therefore of the non
convexity of the rate functional $S$.

\subsection{The rate functional on constant profiles}

Here we show that for constant profiles $\theta$ the
boundary value problem (\ref{Deq}) can be integrated; the
corresponding value of the rate functional $S(\theta)$ can be expressed
in terms of special functions.

We use the variable $\varphi =\log \tau$; we then have
$S(\theta)=\widetilde{\mathcal{G}}(\theta ,\varphi[\theta] )$,
where the functional $\widetilde{\mathcal{G}}$ has been defined in
(\ref{Gtilde}) and $\varphi[\theta]$ is the unique strictly increasing solution
of the boundary value problem
\begin{equation}
\left\{
  \begin{array}{l}
{\displaystyle \vphantom{\Big\{_{B_{B_B}}}
e^{\varphi}
\frac{\varphi''}{(\varphi')^2} + \theta
  = 0 } \\
\varphi(\pm 1) = \log \tau_\pm
  \end{array}
\right.
\end{equation}
If we restrict to constant profiles $\theta$ this equation can be
integrated obtaining
\begin{equation}
\log \varphi'[\theta](u)=\log
\varphi'[\theta](-1)+
\theta\left\{e^{-\varphi[\theta](u)}-\frac{1}{\tau_-}\right\}
\label{Deqf}
\end{equation}
and from this
$$
S(\theta)=\widetilde{\mathcal{G}}(\theta ,\varphi[\theta])
=2
\Big\{-1+\frac{\theta}{\tau_-}-\log \theta -\log
\varphi'[\theta](-1)+\log \frac{{\tau_+-\tau_-}}{2}
\Big\}
$$
{}From equation (\ref{Deqf}) we obtain also
$$
\varphi'[\theta](-1)=\frac 12 e^{\frac{\theta}{\tau_-}}
\int_{\log\tau_-}^{\log \tau_+} \!d\psi \:
e^{-\theta e^{-\psi}}
$$
and finally with a change of variables
\begin{equation}
\label{Sconst}
S(\theta)=2\Big[
-\log\Big( \theta \int_{\frac{1}{\tau_+}}^{\frac{1}{\tau_- }}
\!dy \:
\frac{e^{-\theta y }}{y}
\Big)
-1+\log (\tau_+-\tau_- ) \Big]
\end{equation}
In particular for $\theta$ large,  from (\ref{Sconst}) we deduce the
asymptotic expansion
\begin{equation}
\label{Sconstexp}
S(\theta)=2\Big\{ \frac{\theta}{\tau_+}
+ \Big( \log \frac{\tau_+-\tau_-}{\tau_+} -1 \Big)
+ \frac{\tau_+}{\theta} \Big\}
+ O\Big(\frac{1}{\theta^2}\Big)
\end{equation}

Recall that the equilibrium functional $S_0$ is given in
(\ref{equientr}) and note that for constant and large values of the
profile $\theta$ we have $S_0(\theta)\approx 2 \theta/\tau$.
By comparing this with the expansion (\ref{Sconstexp}),
we see that the leading order is the same but only the warmer
thermostat matters, as it is quite reasonable from a physical point of
view.

As we showed earlier, the rate functional $S$
is not convex. The restriction of $S$ to constant profiles obtained in
(\ref{Sconst}) might however be convex; we do not have an analytic proof of
the convexity of (\ref{Sconst}), but rough numerical evidences
suggest this is the case.

\subsection{An additivity principle}

In \cite{DLS2} the rate functional $S$ was derived for the SEP by
combinatorial techniques.  It was then shown that $S$ satisfies a
suitable \emph{additivity principle} which allows to construct the rate
functional for a system in a macroscopic interval $[a,b]$ from the rate
functional of subsystems in the intervals $[a,c]$ and $[c,b]$, here
$a<c<b$. More precisely, in  \cite{DLS2} is introduced a modified
rate functional $\widetilde S_{[a,b]}(\tau_a,\tau_b;\theta)$ where
$\tau_a$, $\tau_b$ are the density at the endpoints and
$\theta=\theta(u)$ is the density profile in $[a,b]$. The additivity
principle is then formulated as
\begin{equation}
\label{apsep}
\widetilde S_{[a,b]}(\tau_a,\tau_b;\theta)
=\sup_{\tau_c} \;
\big\{ \widetilde S_{[a,c]}(\tau_a,\tau_c;\theta\!\restriction_{[a,c]})
+ \widetilde S_{[c,b]}(\tau_c,\tau_b;\theta\!\restriction_{[c,b]})\big\}
\end{equation}
where $\theta {\displaystyle \!\!\!} \restriction_{[a,c]}$, respectively
$\theta {\displaystyle \!\!\!} \restriction_{[c,b]}$, denotes the
restriction of the profile $\theta$, which is defined on the interval
$[a,b]$, to the subinterval $[a,c]$, respectively $[c,b]$.  The
additivity principle (\ref{apsep}) plays a crucial role in the
derivation of the rate functional for the asymmetric exclusion
process.
In \cite{DLS3} the
expression of the rate functional from this principle is then deduced.

Here we show that the rate functional $S$ for the KMP
process satisfies an additivity principle analogous to (\ref{apsep}).
Here, however, we need to minimize on the midpoint value $\tau_c$. This is
due to the fact that in (\ref{infG}) we need to minimize over the
auxiliary profile $\tau$; a direct derivation of the additivity
formula, as was done in \cite{DLS3} for the asymmetric exclusion
process, would clarify the basic physical difference between the KMP
process and the SEP.

\smallskip
Let us consider the KMP process on the macroscopic interval $[a,b]$,
here we denote the temperatures of the heat baths at the boundary by
$\tau_a$, $\tau_b$. We then let $S_{[a,b]}(\tau_a,\tau_b;\theta)$ be
the corresponding rate functional and introduce
\begin{equation}
\label{tS}
\widetilde S_{[a,b]}(\tau_a,\tau_b;\theta)
= S_{[a,b]}(\tau_a,\tau_b;\theta) - (b-a) \log
\frac{\tau_b-\tau_a}{b-a}
\end{equation}
by using (\ref{funcG}) and (\ref{infG}) we then get
\begin{equation}
\label{tSinfG}
\widetilde S_{[a,b]}(\tau_a,\tau_b;\theta)
= \inf_{\substack{\tau\,:\\ \tau(a)=\tau_a,\,\tau(b)=\tau_b}} \;
\int_{a}^{b} \!du \:
\Big[ \frac{\theta(u)}{\tau(u)} - 1 - \log\frac{\theta(u)}{\tau(u)}
-\log \tau'(u)  \Big]
\end{equation}
where the infimum is over the strictly monotone auxiliary profiles
$\tau(u)$, $u\in[a,b]$.
We then get the additivity principle for the KMP process:
\begin{equation}
\label{apkmp}
\widetilde S_{[a,b]}(\tau_a,\tau_b;\theta)
=\inf_{\tau_c\in [\tau_a,\tau_b]} \;
\big\{ \widetilde S_{[a,c]}(\tau_a,\tau_c;\theta\! \restriction_{[a,c]})
+ \widetilde S_{[c,b]}(\tau_c,\tau_b;\theta\! \restriction_{[c,b]})\big\}
\end{equation}

It is not difficult to show, see \cite{DLS2}, that the expression
(\ref{funcG})--(\ref{infG}) for the rate functional follows from the
additivity rule (\ref{apkmp}).

\bigskip
\subsection {Remarks}

We again conclude  with a few mathematical remarks.  We have
discussed existence and uniqueness of
(\ref{Deq}) only for bounded profiles $\theta$; the extension to
$\theta\in \mc M$ should be however straightforward.  Let $V(\theta)$
be the quasi potential as defined by the variational problem
(\ref{quasipot}). Since the optimal path $\pi^*$ has been explicitly
constructed, the rigorous proof of the upper bound $V(\theta)\le
\inf_{\tau\in\mc T_{\tau_\pm}} \mc G (\theta,\tau)$ should be carried
out as in \cite{BDGJL2}.  The proof of the lower bound $V(\theta)\ge
\inf_{\tau\in {\mc T}_{\tau_\pm}} \mc G (\theta,\tau)$ is instead more
troublesome. The computations presented here essentially prove this
bound for strictly positive smooth paths $\pi$, but the argument in
\cite{BDGJL2} to extend it to arbitrary paths in $\mc E_{\theta,T}$
takes advantages of the convexity (special for the SEP) of the
dynamical rate functional $J$.  This problem is of course related to
the proof of the lower bound for the dynamical large deviation
principle for any open set mentioned at the end of Section \ref{s:db}.
The identification of the quasi potential $V$ with the rate functional
for the invariant measure $S$ has been proven for the SEP in
\cite{BG}; although the strategy is of wider applicability, the
technical points might require some extra effort.

\section{Higher space dimensions}
\label{s:d>1}

The KMP process introduced in Section \ref{secmodel} can be easily
generalized to the case of space dimensions $d>1$.  Let $\Lambda$
be a smooth domain in $\bb R^d$ and set $\Lambda_N:=\bb{Z}^d \cap
N\Lambda$.  We then define the process on $\Sigma_N := \bb
R_+^{\Lambda_N}$as follows:  every pair of nearest neighbors oscillators
exchanges energy according to the rule described in Section
\ref{secmodel} and every oscillator at a boundary site $x$ is in
contact with a thermostat at temperature $\tilde{\tau}(x/N)$ for a
fixed function $\tilde{\tau}$.

Several computations of this paper can be repeated step by step when
the model is not one--dimensional.  In particular the hydrodynamic
equation has still the same form (\ref{he}) with the boundary
condition $\theta {\displaystyle \!\!} \restriction_{\partial \Lambda}
=\tilde{\tau}$ and the dynamic
large deviation functional $J$ has the same form as
(\ref{finalaction}).  {}Formula (\ref{quasipot}) as well as the
Hamilton--Jacobi equation (\ref{hj}) for the quasi potential holds in
any dimension; we can still perform the change of
variables (\ref{cvhj}) and reduce (\ref{hj}) to (\ref{hjspec}).
However the solution of the boundary value (\ref{Deq}) does not give
the quasi potential because, with this choice, the right hand side of
(\ref{cvhj}) is a functional derivative only if $d=1$.

However, by analyzing the variational problem
(\ref{quasipot}), we derive an upper bound for quasi potential
$V(\theta)$ that holds in any space dimension.  We also discuss here
the Gaussian fluctuations of the empirical energy when $\xi$ is
distributed according to the invariant measure. We shall obtain the
covariance of the Gaussian fluctuations by expanding the large
deviations functional $S(\theta)$ around the stationary profile
$\bar\theta$. We note that in the one--dimensional case the arguments
are easier thanks to the more explicit form of $S$.

\subsection{Upper bound for the quasi potential}

Let us denote by $\bar\theta(u)$, $u\in\Lambda$ the stationary
solution of (\ref{he}) with boundary condition
$\bar\theta(u)=\tilde\tau(u)$, $u\in\partial\Lambda$. Note that, for
generic boundary conditions $\tilde\tau$, the profile $\bar\theta$
does not have
the simple form (\ref{bart}). Of course $\bar\theta$ is still
the most likely profile for the empirical energy under the invariant
measure. We also introduce the \emph{local equilibrium} large
deviation function
\begin{equation}
S_{\text{eq}}(\theta) = \int_{\Lambda} \!du \:
\left[ \frac{\theta(u)}{\bar{\theta}(u)} - 1 -
\log\frac{\theta(u)}{\bar{\theta}(u)} \right],
\end{equation}
and note that it coincides with the function defined in
(\ref{finalaction0}) and it is thus the rate functional for the
product measure $\nu_{\bar\theta}^N$.

When $d=1$ we can use (\ref{funcG}) and easily obtain the upper
bound
\begin{equation}
\label{bound}
S(\theta) =\inf_{\tau \in \mathcal{M}_{T_{\tau_\pm}}}
\mathcal{G}(\theta,\tau)
\leq \mathcal{G}(\theta,\bar{\theta})=S_{\text{eq}}(\theta)
\end{equation}

{}For $d>1$ we use a different strategy.  Given a path $\pi=\pi(t,u)$
satisfying the boundary condition $\pi(t,u)=\tilde\tau(u)$,
$u\in\partial\Lambda$, a straightforward computation shows that
\begin{equation*}
\begin{array}{lcl}
{\displaystyle
J_{[-T,0]}(\pi)
}
&= &
{\displaystyle
\frac 12 \int^0_{-T}\!dt \: \Big\langle
\nabla^{-1} \Big[ \partial_t \pi -\frac 12 \Delta \pi \Big] ,
\frac{1}{\pi^2}
\nabla^{-1} \Big[ \partial_t \pi -\frac 12 \Delta \pi \Big]
\Big\rangle
}
\\
&= &
{\displaystyle \vphantom{\Bigg\{^K}
\frac 12
\int^0_{-T}\!dt \: \Big\langle
\nabla^{-1}\Big[ \partial_t \pi + \frac 12 \Delta \pi
+\nabla\Big(\pi^2 \nabla\frac{1}{\bar\theta}\Big)
-\Delta \pi - \nabla\Big(\pi^2 \nabla\frac{1}{\bar\theta}\Big)
\Big] ,
}
\\
&&
{\displaystyle
\phantom{\frac 12 \int^0_{-T}\!dt} \vphantom{\bigg\{}
\frac{1}{\pi^2}
\nabla^{-1}\Big[ \partial_t \pi + \frac 12 \Delta \pi
+\nabla\Big(\pi^2 \nabla\frac{1}{\bar\theta}\Big)
-\Delta \pi - \nabla\Big(\pi^2 \nabla\frac{1}{\bar\theta}\Big)
\Big]
\Big\rangle
}
\\
&= &
{\displaystyle \vphantom{\Bigg\{}
S_{\text{eq}}(\pi(0))-S_{\text{eq}}(\pi(-T))
- \frac 12 \int^0_{-T} \! dt\:
\Big\langle \frac{(\nabla \bar{\theta})^2}{\bar{\theta}^4},
(\pi - \bar{\theta})^2\Big\rangle
}
\\
&&
{\displaystyle
+\:\frac 12
\int^0_{-T}\! dt \:
\Big\langle
\nabla^{-1}
\Big[ \partial_t \pi +\frac 12 \Delta \pi
+ \nabla\Big( \pi^2 \nabla \frac{1}{\bar{\theta}} \Big)
\Big],
}
\\
&&
{\displaystyle
\phantom{ \int^0_{-T}\! dt \:  \Big\langle }
\quad
\frac{1}{\pi^2}
\nabla^{-1}
\Big[ \partial_t \pi +\frac 12 \Delta \pi
+ \nabla\Big( \pi^2 \nabla \frac{1}{\bar{\theta}} \Big)
\Big] \Big\rangle
}
\end{array}
\end{equation*}
The quasi potential is defined by the variational
problem (\ref{quasipot}).  Hence,
to obtain an upper bound for $V(\theta)$ it is enough to
exhibit a path $\pi$ which connects $\bar\theta$ to $\theta$. We
choose $\pi(t)=\tilde \theta(-t)$ where $\tilde\theta(t)$ solves
\begin{equation*}
\left\{
\begin{array}{lcl}
{\displaystyle
\partial_t \tilde\theta(t)
}
&=&
{\displaystyle
\frac 12 \Delta \tilde\theta(t)
+ \nabla \Big( \tilde{\theta}(t)^2 \nabla \frac{1}{\bar{\theta}} \Big)
}
\\
{\displaystyle \vphantom{\bigg\{}
\tilde\theta (t, u)
}
&=&  {\displaystyle  \tilde\tau(u) }
\,,\quad u\in\partial\Lambda
 \\
{\displaystyle \tilde\theta(0,u)}
&=&
{\displaystyle \theta(u)}
\end{array}
\right.
\end{equation*}
We note that the path $\pi$ connects $\bar\theta$ to $\theta$ since
$\tilde\theta(t)\to\bar\theta$ as $t\to\infty$.
By using the path $\pi$ in the above expression for
$J_{[-T,0]}(\pi)$ and letting $T\to\infty$ we get
\begin{equation*}
V(\theta)
\le S_{\text{eq}}(\theta) -
\frac 12 \int^0_{-\infty} \! dt \:
\Big \langle \frac{(\nabla \bar{\theta})^2}{\bar{\theta}^4},
[\pi(t) - \bar{\theta}]^2\Big\rangle
\leq S_{\text{eq}}(\theta)
\end{equation*}
which shows that the upper bound $V(\theta) \le S_{\text{eq}}(\theta)$
holds in any space dimension.  We also note that the above inequality
is strict unless $\theta=\bar\theta$.

\subsection{Gaussian fluctuations}

In the case $d=1$ we can follow step by step the argument given in
\cite{DLS2} for the SEP.  We consider a small perturbation,
$\theta=\bar{\theta}+\epsilon\Theta$, of the stationary profile
$\bar{\theta}$, and consequently have
$\tau[\theta]=\bar{\theta}+\epsilon T$, where, to first
order in $\epsilon$, (\ref{Deq}) gives
\begin{equation}
\frac{\bar{\theta}^2}{(\nabla \bar{\theta})^2}\Delta T -T
=-\Theta
\label{primordineDeq}
\end{equation}
The functional $S(\theta)$ has a minimum at $\bar{\theta}$ so that its
expansion in $\epsilon$ is
\begin{equation}
S(\theta )=S(\bar{\theta})+
\frac{1}{2}\epsilon^2 \langle\Theta,C^{-1} \Theta\rangle
+o\left(\epsilon^2\right)
\label{expS}
\end{equation}
The operator $C$ is the covariance for the Gaussian fluctuations
of the empirical energy under the invariant measure $\mu_{N,\tau_\pm}$.
Since $S(\theta)=\mc G (\theta,\tau[\theta])$, we get
\begin{equation*}
\begin{array}{lcl}
{\displaystyle
\langle\Theta,C^{-1}\Theta\rangle
}
& = &
{\displaystyle
\int_{-1}^1\! du \:
\Big\{ \frac{[ T (u)-\Theta(u)]^2}{\bar{\theta}^2(u)}
+ \frac{[ \nabla T (u )] ^2 }{[ \nabla\bar{\theta}]^2}
\Big\}
}
\\
& = &
{\displaystyle \vphantom{\Bigg\{^{K}}
\int_{-1}^1\! du \:
\Big\{
\frac{ \bar{\theta}(u)^2 [ \Delta T(u)]^2}
{[ \nabla\bar{\theta}]^4}
- \frac{T(u) \Delta T(u)}{[\nabla\bar{\theta}]^2}
\Big\}
= \Big\langle \Delta T, \frac{W}{(\nabla\bar{\theta})^2}
\Delta T \Big\rangle
}
\end{array}
\end{equation*}
where we used Taylor expansions, integrations by parts, $T(\pm 1) =0$,
and (\ref{primordineDeq}).  The operator $W$ is defined as
\begin{equation}
W:=\frac{\bar{\theta}^2}{(\nabla\bar{\theta})^2}\id + (-\Delta)^{-1}
\end{equation}
{}From equation (\ref{primordineDeq}) we get
$\Theta= - W \Delta T$ and this implies
\begin{equation*}
\langle\Theta,C^{-1}\Theta\rangle
= \Big\langle\Delta T ,
\frac{W}{(\nabla\bar{\theta})^2}\Delta T \Big\rangle
= \Big\langle \Theta, \frac{W^{-1}}{(\nabla\bar{\theta})^2}
\Theta\Big\rangle
\end{equation*}
so that the covariance $C$ is given by
\begin{equation}
\label{Cd=1}
C= (\nabla\bar{\theta})^2 \, W
=\bar{\theta}^2 \id + (\nabla\bar{\theta})^2 \, (-\Delta)^{-1}
\end{equation}
The first term above is simply the covariance of the Gaussian
fluctuations of the empirical energy for local equilibrium product
measure $\nu_{\bar\theta}^N$, while the second term
represents the contribution to the covariance due to the long range
correlations in the stationary nonequilibrium state.
As in the case of the SEP \cite{BDGJL1,DFIP,DLS1,DLS2,S} this correction
is given by $(-\Delta)^{-1}$, the Green function of the Dirichlet
Laplacian.
Since $(-\Delta)^{-1} > 0$, for the KMP process this correction
enhances the Gaussian fluctuations, while in the case of SEP it
decreases them.
We also note that, by exploiting the duality of the KMP process with
the process we shall introduce in Section \ref{s:du}, the expression
(\ref{Cd=1}) for the covariance could be rigorously deduced as in the
case of the SEP \cite{DFIP,S}.

\medskip
To obtain the covariance of Gaussian fluctuations in the case
$d>1$ we instead argue as in \cite{BDGJL1}.
Let us introduce the ``pressure'' as the Legendre transform of the
rate functional $S(\theta)$, i.e.\
\begin{equation}
G(h):=\sup_{\theta}\big\{ \langle \theta,h \rangle -S(\theta )\big\}
\label{legge}
\end{equation}
We then get that $G(h)$ satisfies the Hamilton--Jacobi equation dual
to (\ref{hj}), i.e.\
\begin{equation}
\label{HJP}
\Big\langle \nabla h,
\Big( \frac{ \delta G}{\delta h} \Big)^2
\nabla h \Big\rangle
=\Big\langle \nabla h , \nabla  \frac{ \delta G}{\delta h} \Big\rangle
\end{equation}
where $h(u)$ satisfies the boundary conditions
$h(u)|_{\partial\Lambda}=0$.

Let us denote by $G_{\text{eq}}$ the pressure associated via (\ref{legge}) to
the local equilibrium functional $S_{\text{eq}}$;
we look for a solution of (\ref{HJP}) in the form
\begin{equation}
G(h)=G_{\text{eq}}(h)+
\langle g,h \rangle +\frac{1}{2}
\langle h,B \,h\rangle + o(h^2)
\label{G=}
\end{equation}
for some function $g=g(u)$ and some linear operator $B$. {}From
(\ref{expS}) we get
\begin{equation}\label{G2}
G(h) = \langle \bar\theta, h \rangle
+\frac 12 \left\langle h, C \, h \right\rangle + o(h^2)
\end{equation}
hence the second derivative of $G$ at $h=0$ is the covariance of the
density fluctuations. By comparing (\ref{G=}) to (\ref{G2}) we find
\begin{equation}
\label{Cov}
C = \frac {\delta^2  G_{\text{eq}} }{\delta h^2}\Big|_{h=0} + B
= \bar{\theta}^2 \, \id  + B
\end{equation}
By plugging (\ref{G=}) into (\ref{HJP}) and expanding up to
second order in $h$, it is not difficult to show,
see \cite{BDGJL1} for the case of the SEP, that $g=0$ and
\begin{equation}
\langle h, \Delta B \, h \rangle =
- \big\langle h, |\nabla\bar{\theta} |^2 h \big\rangle
\end{equation}
The operator $B$ therefore satisfies
\begin{equation}
\label{eqB}
\frac 12 [ \Delta B + B \Delta ]
= -|\nabla \bar{\theta}|^2
\end{equation}
See \cite{S} for another derivation of this equation based on
fluctuating hydrodynamic instead of large deviations.

{}From (\ref{eqB}) we see that if $\nabla\bar\theta$ is constant
(this condition can be realized by a suitable choice of the
thermostat $\tilde\tau$), the operator $B$ has the kernel
\begin{equation}
\label{kerB}
B(u,v) = |\nabla\bar\theta|^2 (-\Delta)^{-1}(u,v)
\end{equation}
where $(-\Delta)^{-1}(u,v)$ is the Green function of the Dirichlet
Laplacian in $\Lambda$.  The interpretation of (\ref{Cov}) and
(\ref{kerB}) is as in the one dimensional case (\ref{Cd=1}); we note
the fact that $B$ is a positive operator can also be obtained as a
consequence of the bound $S(\theta) \leq S_{\text{eq}} (\theta)$.

\section{The dual process}
\label{s:du}

The analysis in \cite{KMP} is based on a duality relationship between
the KMP process and another process we discuss next in the
one--dimensional case. The state space is $\Sigma_N := \bb
N^{\Lambda_N}$ where $\bb N :=\{0,1,\dots\}$ is the set of natural
numbers. If $\xi=\{\xi_x\,,\: x\in\Lambda_N\}\in \Sigma_N$, the value
$\xi_x$ at the site $x$ can therefore be interpreted as the number of
particles at $x$. As for the KMP process, at each bond $\{x,x+1\}$ there
is an exponential clock of rate one; when it rings the total number of
particles $\xi_x+\xi_{x+1}$ is redistributed uniformly across the bond
$\{x,x+1\}$. Moreover the boundary sites $\pm N$ evolve according to a
heat bath 
dynamics with respect to a geometric distribution with parameter
$p_\pm$.  More formally, the infinitesimal generator has still the form
(\ref{gen}) but now the bulk dynamics is defined by
\begin{equation}
\label{genD}
L_{x,x+1}f \, (\xi)
:=\frac{1}{\xi_x+\xi_{x+1}+1}
\sum_{k=0}^{\xi_x+\xi_{x+1}}
\big[ f(\xi^{(x,x+1),k})-f(\xi) \big]
\end{equation}
where the configuration $\xi^{(x,x+1),k}$ is defined as
\begin{equation}
\big( \xi^{(x,x+1),k} \big)_y
:=\left\{
\begin{array}{ll}
\xi_y & \textrm{ if } \ \ y\neq x,x+1 \\
k & \textrm{ if  }\ \ y=x \\
\xi_x+\xi_{x+1}-k & \textrm{ if } \ \ y=x+1
\end{array}
\right.
\nonumber
\end{equation}
The boundary part of the generator is defined as follows
\begin{equation}
L_{\pm}f \, (\xi ):=\sum_{k=0}^{\infty}p_{\pm}(1-p_{\pm})^k
\big[ f(\xi^{\pm N,k})-f(\xi )\big]
\end{equation}
where $p_\pm \in (0,1)$ are the parameters of the reservoirs and
the configuration $\xi^{x,k}$ is defined as
\begin{equation}
\big( \xi^{x,k} \big)_y
:=\left\{
\begin{array}{ll}
\xi_y  & \textrm{ if } \ \ y\neq x  \\
k & \textrm{ if } \ \ y=x \\
\end{array}
\right.
\end{equation}

If $p_+=p_-=p$ the dynamics is reversible with respect to the
product of geometric distributions of parameter $p$, i.e.\ the
invariant measure is
\begin{equation}
\mu_{N,p}(\xi)= \prod_{x\in\Lambda_N}
\big[ p\, (1-p)^{\xi_x}\big]
\end{equation}
By a computation analogous to (\ref{ldpr})--(\ref{press}), it is easy
to show that when $\xi$ is distributed according to $\mu_{N,p}$ then
the empirical density $\pi_N(\xi)$, which is defined as in
(\ref{eme}), satisfies a large deviation principle with the rate
functional
\begin{equation}
S_0(\theta )=\int_{-1}^1 \! du\:
\Big\{
\theta(u)\log\frac{\theta(u)}{\bar{\theta}}+[1+\theta(u)]
\log\frac{1+\bar{\theta}}{1+\theta(u)}
\Big\}
\end{equation}
where the parameter $\bar{\theta}$ is related to $p$ by the relation
$\bar{\theta}=\sum_{k=0}^{\infty}k\,p(1-p)^k=(1-p)/{p}$.

When $p_+\neq p_-$ the model is no longer reversible and the
invariant measure $\mu_{N,p_\pm}$ is not explicitly known. In the
sequel we shall assume $p_->p_+$.
We can repeat the computations done for the KMP process and get the
hydrodynamic equation.  This is still the linear heat equation with the
appropriate boundary conditions, i.e.\
\begin{equation}
\label{heD}
\left\{
\begin{array}{lcl}
{\displaystyle
\partial_t \theta(t)
}
&=& {\displaystyle \frac 12 \Delta \theta(t)
}\\
{\displaystyle \vphantom{\bigg\{}
\theta(t, \pm1)}
&=&  {\displaystyle  \theta_{\pm} \:=\: \frac {1-p_\pm}{p_\pm}
}\\
{\displaystyle \theta(0,u)}
&=&
{\displaystyle \theta_0(u)}
\end{array}
\right.
\end{equation}
As before the most likely density profile $\bar\theta$ is the
stationary solution of (\ref{heD}).

To obtain the dynamic large deviation principle we introduce a smooth
function $H= H(t,u)$ vanishing at the boundary and consider the following
time dependent perturbation of the generator $L_{x,x+1}$ in (\ref{genD})
\begin{equation*}
\begin{array}{lcl}
{\displaystyle L_{x,x+1}^{H}f \, (\xi) } &:=& {\displaystyle
\frac{1}{\xi_x+\xi_{x+1}+1}\sum_{k=0}^{\xi_x+\xi_{x+1}}
e^{(\xi_x-k)[ H(t,(x+1)/{N})-H(t,{x}/{N})]}
}\\
&&
{\displaystyle
\phantom{\frac{1}{\xi_x+\xi_{x+1}+1}\sum_{k=0}^{\xi_x+\xi_{x+1}}}
\times \:
\big[ f( \xi^{(x,x+1),k} ) - f ( \xi ) \big]
}
\end{array}
\end{equation*}
The hydrodynamic equation associated to this perturbed dynamics is
given by
\begin{equation}
\label{idroFD}
\partial_t \theta(t) =
\frac 12 \Delta \theta(t) - \nabla\Big( \theta(t)
[ 1+\theta(t) ]  \nabla H(t) \Big)
\end{equation}
with the same boundary conditions as (\ref{heD}).
By the same computations as in Section \ref{s:db},
we get that the dynamical large deviation functional is
\begin{equation}
\label{finalactionD}
J_{[0,T]}(\pi ) =  \frac 12 \int_0^T\!dt
\: \big\langle \nabla H(t) , \pi(t) [ 1+\pi(t)]
\nabla H(t) \big\rangle
\end{equation}
where $H$ has to be obtained from the path $\pi$ by using equation
(\ref{idroFD}) with $\theta(t)$ replaced by $\pi(t)$.

This leads to the following Hamilton--Jacobi equation for the
quasi potential
\begin{equation}
\label{hjD}
\Big\langle \nabla \frac{\delta V}{\delta \theta},
\theta(1+\theta) \nabla \frac{\delta V}{\delta \theta} \Big\rangle
+\Big\langle \frac{\delta V}{\delta \theta}, \Delta \theta
\Big\rangle =0
\end{equation}
where ${\delta V}/{\delta \theta}$ vanishes at the boundary
and $\theta(\pm 1)=\theta_\pm$.

We look for a solution of the form
\begin{equation}
\frac{\delta V}{\delta \theta}=\log \frac{\theta}{1+\theta}-\log
\frac{F}{1+F}
\end{equation}
By the same computations as in Section \ref{s:qp}, we reduce
the Hamilton--Jacobi equation (\ref{hjD}) to
\begin{equation}
\Big\langle \frac{\theta -F}{F^2(1+F)^2}\,,\,
F(1+F)\Delta F+(\theta -F)(\nabla F)^2 \Big\rangle=0
\end{equation}
We thus obtain a solution of (\ref{hjD}) considering the functional
\begin{equation}
\label{FD}
V(\theta )=
\int_{-1}^1 du\:
\Big\{
\theta(u)\log\frac{\theta(u)}{F(u)}+[1+\theta(u)]
\log\frac{1+F(u)}{1+\theta(u)}-
\log \frac{F'(u)}{[\theta_+-\theta_-]/2}
\Big\}
\end{equation}
where $F(u)$ has to be computed from $\theta(u)$ as the unique
strictly increasing solution of the boundary value problem
\begin{equation}\label{Deq2D}
\left\{
\begin{array}{l}
{\displaystyle
F(1+F) \frac{F''}{\big( F' \big)^2} +
\theta -F
  = 0 } \\
{\displaystyle \vphantom{\Big\{}
F(\pm 1) = \theta_\pm
}
  \end{array}
\right.
\end{equation}
As for the KMP process it is possible to check that
this is the right solution of the Hamilton--Jacobi equation (\ref{hjD}).

By the change of variable $F=e^{\varphi}$, it is easy to verify  that
the right hand side of (\ref{FD}) is strictly convex in $\varphi$.
We therefore have, analogously to the KMP process,
\begin{equation*}
V(\theta) =\inf_{F}\;
\int_{-1}^1 du\:
\Big\{
\theta(u)\log\frac{\theta(u)}{F(u)}+[1+\theta(u)]
\log\frac{1+F(u)}{1+\theta(u)}-
\log \frac{F'(u)}{[\theta_+-\theta_-]/2}
\Big\}
\end{equation*}
where the infimum is carried out over all strictly increasing
functions $F$ satisfying the boundary condition $F(\pm 1)=\theta_\pm$.

\section{Conclusions: few comments on generic models}
\label{s:7}

{}For the SEP, the derivation of the rate function for the stationary non
equilibrium state obtained in \cite{DLS1,DLS2}  depends
heavily on the details of the microscopic process. On the other
hand, the variational approach in \cite{BDGJL1} depends only on the
macroscopic transport coefficients, bulk diffusion $D$ and mobility
$\sigma$
of the system.
These are not
independent functions,  they are related by the Einstein
relation $D(\rho) = \sigma(\rho) \chi(\rho)^{-1}$ where
$\chi(\rho)$ is the \emph{compressibility}. It is defined as
$\chi(\rho)^{-1} = \lambda'(\rho)= f_0''(\rho)$ where $f_0$
is the (equilibrium) Helmholtz free energy of the system and
$\lambda$ is the chemical potential.
This means in particular that while the derivation in
\cite{DLS1,DLS2} is only valid for nearest neighbor jumps, the result
holds for the general SEP.
 In this paper we have
discussed a model, the KMP process (in fact two models if we
consider also its dual process), in which the rate functional has
an expression very similar to the one for the SEP. Here we discuss
what are the essential features of the functional form of these
coefficients  in the derivation of the rate
functional $S$.  In this discussion
we shall consider $D$ and $\sigma$ as given and discuss the large
deviations properties of the nonequilibrium state.

We discuss only one--dimensional (symmetric) diffusive system with a single
conservation law and particle reservoirs at the boundary. Here it will be
convenient to think of the conserved quantity as the density of
particles. {}For general models, the hydrodynamic equation is expected
(proven for equilibrium models in \cite{VY} and in \cite{ELS1,ELS2}
for nonequilibrium under the so called \emph{gradient condition}),
to be given by a nonlinear diffusion equation with Dirichlet data at
boundary, i.e.\
\begin{equation}
\label{7.1}
\left\{
\begin{array}{lcl}
{\displaystyle
\partial_t \rho(t,u)}
&=& {\displaystyle \frac 12 \nabla
\Big( D\big(\rho(t,u)\big) \nabla \rho(t,u)\Big) }\\
{\displaystyle \vphantom{\bigg\{}
\rho(t, \pm1)}
&=&  {\displaystyle  \rho_\pm } \\
{\displaystyle \rho(0,u)}
&=&
{\displaystyle \rho_0(u)}
\end{array}
\right.
\end{equation}
where the bulk diffusion $D(\rho)= \sigma(\rho) \chi(\rho)^{-1}$
is given by a Green--Kubo formula, see e.g.\ \cite[II.2.2]{Slib}.
For the KMP process, as well as for the SEP, we simply have $D=1$,
i.e.\ $\sigma = \chi$.

The probability of a large deviations from the hydrodynamic behavior
are expected (to our knowledge for open systems this has been proven
only for the SEP in \cite{BDGJL2}, see however \cite[II.3.7]{Slib} for
an heuristic derivation for equilibrium lattice gas models) to have the form
(\ref{dldp})--(\ref{dldpbis}) where the dynamical cost $J_{[0,T]}$
should be of the form
\begin{equation}
\label{finalactionbis}
J_{[0,T]}(\pi) = \frac 12 \int_0^T\!dt \:
\langle \nabla H (t), \sigma(\pi(t))\nabla H(t) \rangle
\end{equation}
in which the perturbation $H$ has to be chosen so that the
fluctuation $\pi$ solves the perturbed hydrodynamic
\begin{equation}
\label{idroFbis}
\left\{
\begin{array}{lcl}
{\displaystyle
\partial_t \pi (t)
}
&=&
{\displaystyle
\frac 12 \nabla \Big( D\big(\pi(t) \big) \nabla \pi (t) \Big)
- \nabla \Big( \sigma\big(\pi(t)\big)  \nabla H(t) \Big)
}
\\
{\displaystyle \vphantom{\bigg\{} \pi(t, \pm 1)
}
&=&
{\displaystyle
\rho_\pm
}
\\
{\displaystyle \pi(0,u)
}
&=&
{\displaystyle
\rho_0(u)
}
\end{array}
\right.
\end{equation}
and $\sigma(\pi)$ is the \emph{mobility} of the system.
{}For the SEP process we have $\sigma(\pi)= \pi (1-\pi)$ (note that in
this case we have $0\le \pi\le 1$) while for the KMP process,
respectively its dual, we have $\sigma(\pi)=\pi^2$, respectively
$\sigma(\pi)= \pi (1+\pi)$.

\medskip
We first mention the few examples in which it is possible to obtain the
rate function $S$ in a closed form. The following models are however
even simpler than the SEP or the KMP process since they do not exhibit
the non--locality of $S$, which reflects, at the large deviation
level, the long range correlations of the system which are expected
\cite{BJ,S} to be a generic feature of nonequilibrium models.

The easiest example is provided by independent particles.  In this
case we have $D$ constant and $\sigma$ linear.  The nonequilibrium
state is a product measure and it is easy to verify that
$S(\rho)=\int_{-1}^{1}\!du\: f\big(\rho(u),\bar\rho(u)\big)$, where
\begin{equation}
\label{frt}
f(\rho,\tau)= f_0(\rho) - f_0(\tau) + (\rho - \tau)f^\prime_0(\tau)
= \rho \log (\rho/\tau) - (\rho-\tau)
\end{equation}
and $\bar\rho$ is the stationary solution of (\ref{7.1}).
Another example is the so called zero range process, see e.g.\
\cite[II.7.1]{Slib}. In this case $D(\rho)=\Phi'(\rho)$ and
$\sigma(\rho)=\Phi(\rho)$ where the (increasing) function $\Phi$
depends on the microscopic rates. As shown in \cite{DF} the
nonequilibrium state is again a product measure and, as discussed in
\cite{BDGJL3,BDGJL1}, its rate function is
$S(\rho)=\int_{-1}^{1}\!du\: f\big(\rho(u),\bar\rho(u)\big)$ for
$f$ again given by (\ref{frt}) with the appropriate $f_0$. .
These examples (the first being a special case of the second) are
characterized by the fact that $\sigma(\rho) = C
\exp\{\lambda(\rho)\}$ where $C>0$ is a constant and
$\lambda$ is the chemical potential.
The Einstein formula then gives $D(\rho) = \sigma'(\rho)$.
The last example is the Ginzburg--Landau model, see e.g.\
\cite[II.7.3]{Slib}, where $\sigma$ is a constant while $D$
is determined by the Einstein relation.  In this case the
nonequilibrium state is still a product 
measure and its rate function has the same expression as in the zero
range process.

\smallskip
We note that for the SEP, as well as for the KMP process and its
dual, we have $D(\rho)$ constant and $\sigma(\rho)$
a second order polynomial in $\rho$.
We next show that an expression of the rate function $S$ of the
nonequilibrium state can be derived under a general hypothesis.
More precisely, we shall assume that the diffusion coefficient
$D(\rho)$ and the mobility $\sigma(\rho)$ satisfy the following
condition.  There exists a constant $a\in \bb R$ such that for any
$\rho\neq \tau$
\begin{equation}
\label{7.2}
\frac{\sigma(\rho) - \sigma(\tau)}{\int_\tau^\rho \!dr \: D(r)}
= \frac{\sigma'(\tau)}{D(\tau)} + a \, (\rho -\tau)
\end{equation}
This condition, of course, identifies a rather tiny class of
models that in fact coincides with the class of all the examples
discussed. We should not expect to be able to obtain $S$ in almost
a closed form for \emph{any} model.
As we shall see, the locality of the functional $S$ corresponds to
the special case (in this class) $a=0$.

Let us first discuss which functions $D$ and $\sigma$ satisfy
condition (\ref{7.2}). We rewrite it with $\rho$ and $\tau$
exchanged
$$
\frac{\sigma(\tau)-\sigma(\rho)}{\int_{\rho}^{\tau}dr\
D(r)}=\frac{\sigma'(\rho)}{D(\rho)}+a(\tau-\rho)
$$
This equation together with (\ref{7.2}) imply
$$
\frac{\sigma'(\tau)}{D(\tau)}-\frac{\sigma'(\rho)}{D(\rho)}=2a(\tau-\rho)
$$
It is easy to see that this is equivalent to
\begin{equation}
\label{necess1}
\frac{\sigma'(r)}{D(r)}=2ar+c
\end{equation}
with $c$ an arbitrary constant. Condition (\ref{necess1}) is a
necessary condition for the validity of (\ref{7.2}). We rewrite
(\ref{necess1}) in the integrated form
$$
\sigma(\rho)-\sigma(\tau)
=2a\int_{\tau}^{\rho}\!dr \: r D(r)
+ c \int_{\tau}^{\rho}\!dr \; D(r)
$$
and substitute it inside (\ref{7.2}). We thus obtain
\begin{equation}
\label{suff1}
\frac{2a \int^{\rho}_{\tau} \! dr\: rD(r)}
{\int^{\rho}_{\tau}\!dr \: D(r)}= a(\rho+\tau)
\end{equation}
A pair $(\sigma,D)$ is a solution of (\ref{7.2}) if and only if is
a solution of (\ref{necess1}) and (\ref{suff1}).

When $a=0$, equation (\ref{suff1}) is always satisfied and
(\ref{necess1}) becomes $\sigma'(r)=cD(r)$. If $c\neq 0 $ we have the
solutions corresponding to zero range dynamics (with an extra
multiplicative factor $c$); if $c=0$ we have the solutions
corresponding to Ginzburg--Landau models.

When $a\neq 0$, equation (\ref{suff1}) becomes
\begin{equation}
2\int^{\rho}_{\tau}\! dr \: rD(r)
=(\rho+\tau)\int^{\rho}_{\tau}\!dr\: D(r)
\label{suff2}
\end{equation}
We differenciate with respect to $\rho$ and obtain
$$
(\rho-\tau)D(\rho)=\int^{\rho}_{\tau}\!dr\: D(r)
$$
that is satisfied if and only if $D$ is constant. Now condition
(\ref{necess1}) imposes that $\sigma(\rho)$ is a second order
polynomial in $\rho$. In this class of solutions fall the simple
exclusion model, the KMP model and its dual.


\smallskip
To write the rate functional $S$ we need to introduce a little more
notation. We let $d(\rho)=\int_0^\rho\!dr \: D(r)$, since $D>0$ the
function $d$ is strictly increasing; we denote its inverse by $b$.
We finally set $A(\varphi):= \sigma\big(b(\varphi)\big)$.
We next denote partial derivatives by a subscript.
Let us introduce a function of two variables $f=f(\rho,\tau)$ such
that $f_{\rho\rho}(\rho,\tau)= D(\rho) \sigma(\rho)^{-1}=\chi^{-1}(\rho)$  and
normalize $f$ so that $f(\cdot,\tau)$ has a minimum at $\tau$ and
$f(\tau,\tau)=0$. Therefore
$$
f(\rho,\tau)= \int_\tau^\rho \!dr \int_\tau^r \! dr'
\: \frac{1}{\chi(r^\prime)}
$$
It is easy to verify that in the
equilibrium case, $\rho_+=\rho_-=\bar\rho_0$,
the rate function $S_0$ is simply given by
$S_0(\rho)=\int_{-1}^1\!du\: f\big(\rho(u),\bar\rho_0\big)$.
To obtain the rate function $S$ in the nonequilibrium case
$\rho_+\neq\rho_-$ we introduce the functional of two variables
\begin{equation}
\label{Gbis} \mc G (\rho,\varphi) := \int_{-1}^1\!du\: \Big\{
f\big(\rho(u),b(\varphi(u))\big) - \frac{1}{a} \log \frac{\nabla
\varphi(u)}{\nabla d(\bar\rho(u))} \Big\}
\end{equation}
where $\bar\rho$ is the equilibrium profile. Note that
$\nabla d(\bar\rho(u)) = D(\bar \rho(u)) \nabla u$ is a constant since
its divergence must vanish in the stationary state.

We claim that, under condition (\ref{7.2}),
the rate function $S$ can be expressed as
$S(\rho)= \mc G (\rho,\varphi[\rho])$ where, given $\rho$, the
auxiliary function $\varphi=\varphi[\rho]$
is the solution of the Euler--Lagrange equation
$\delta\mc G (\rho,\varphi) / \delta\varphi = 0$, that is
\begin{equation}\label{EL}
\left\{
  \begin{array}{l}
{\displaystyle \vphantom{\Big\{_{B}}
\frac{1}{a} \frac{\Delta \varphi}{\big(\nabla\varphi \big)^2}
+ \frac{\rho - b(\varphi)}{A(\varphi)}
  = 0 } \\
{\displaystyle \vphantom{\Big\{^{B}}
d(\varphi(\pm 1)) = d(\rho_\pm)
}
  \end{array}
\right.
\end{equation}
where we used
$f_\tau(\rho,\tau)= - D(\tau) (\rho - \tau)/\sigma(\tau)$,
$b'(\varphi)= D(b(\varphi))^{-1}$ and the definition of $A$.

The definition of the functional $\mc G$ and the above equation are
not really meaningful if $a=0$, as it is the case for the simple
models in which $S$ is local discussed above. However, in such a case
we understand (\ref{EL}) as $\Delta \varphi =0$
whose solution is $\varphi(u)= d(\bar\rho(u))$.
Plugging it into the functional $\mc G$
we get, by understanding $(\log 1)/ 0 =0$, the correct local functional
$S(\rho)$. On the other hand, as soon as $a\neq 0$, the functional $S$
is nonlocal.

To establish the claim, we next show that the functional $S$
solves the Hamilton--Jacobi equation
\begin{equation}
\label{hjbis}
\Big\langle \nabla \frac{\delta S}{\delta \rho}, \sigma(\rho)
\nabla \frac{\delta S}{\delta \rho} \Big\rangle
+\Big\langle  \frac{\delta S}{\delta \rho},
\nabla\Big( D(\rho) \nabla \rho\Big)
\Big\rangle =0
\end{equation}
The argument to conclude the identification of $S$ with the quasi
potential, as defined in (\ref{quasipot}), is indeed essentially the one
carried out in section \ref{s:ep} and it is therefore omitted.

By the definition of $S$, we get
$\delta S(\rho) / \delta \rho =
\delta \mc G (\rho,\varphi) / \delta \rho =
f_\rho (\rho,b(\varphi)) $ so that
the left hand side of
(\ref{hjbis}), after an integration by parts, reduces to
$$
\begin{array}{l}
{\displaystyle
\Big\langle
f_{\rho\rho}(\rho,b(\varphi)) \nabla\rho
+ \frac{f_{\rho\tau}(\rho,b(\varphi))}{D(b(\varphi))}
\nabla\varphi ,
}
\\
{\displaystyle \vphantom{\bigg\{}
\quad\quad\quad\quad\quad\quad
\sigma(\rho) f_{\rho\rho}(\rho,b(\varphi))\nabla\rho
+ \sigma(\rho) \frac{f_{\rho\tau}(\rho,b(\varphi))}{D(b(\varphi))}
\nabla\varphi
- D(\rho) \nabla\rho
\Big\rangle
}
\\
\quad\quad
{\displaystyle \vphantom{\Bigg\{}
=
\Big\langle
f_{\rho\rho}(\rho,b(\varphi)) \nabla\rho
+ \frac{f_{\rho\tau}(\rho,b(\varphi))}{D(b(\varphi))}\nabla\varphi ,
\sigma(\rho)
\frac{f_{\rho\tau}(\rho,b(\varphi))}{D(b(\varphi))}
\nabla\varphi
\Big\rangle
}
\\
\quad\quad
{\displaystyle =
\Big\langle \nabla d(\rho),
\frac{-\nabla\varphi }{A(\varphi)}
\Big\rangle
+ \Big\langle \nabla\varphi ,
\frac{\sigma(\rho)}{A(\varphi)^2}\nabla\varphi \Big\rangle
}
\end{array}
$$
where we used the Einstein relation
$\sigma(\rho) f_{\rho\rho}(\rho,b(\varphi)) = D(\rho)$ in the first step and
$f_{\rho\tau}\big(\rho,b(\varphi)\big)= - D(b(\varphi))/A(\varphi)$ in
the second one.

We next write $\nabla d(\rho)=\nabla [d(\rho)-\varphi]+\nabla\varphi$ and
integrate by parts the first term in the last expression above (recall
that $d(\rho)$ and $\varphi$ satisfy the same boundary conditions).
We finally get that the left hand side of (\ref{hjbis}) equals
$$
\bigg\langle d(\rho) -\varphi ,
\nabla \Big( \frac{\nabla\varphi}{A(\varphi)} \Big)
+ \frac{\sigma(\rho) - A(\varphi)}{d(\rho) -\varphi}
\frac{1}{A(\varphi)^2} \big(\nabla\varphi\big)^2 \bigg\rangle
$$
We therefore find that the functional $S$ solves the
Hamilton--Jacobi equation (\ref{hjbis}) provided $\varphi$ satisfies
the equation
\begin{equation}
\label{Deqbis}
A(\varphi) \Delta \varphi
+\Big[ - A'(\varphi)+ \frac{\sigma(\rho) - A(\varphi)}{d(\rho) -\varphi}\Big]
\big(\nabla\varphi\big)^2 = 0
\end{equation}
In general, we have no reason to expect to be able to express
the solution of the functional derivative equation
(\ref{hjbis}) by a boundary value problem analogous to (\ref{Deqbis}),
it simply works under our special assumption.

Up to this point we did not yet really use condition (\ref{7.2})
but, to complete the argument, we need to show that (\ref{Deqbis})
is equivalent to the Euler--Lagrange equation (\ref{EL}).
By writing (\ref{7.2}) with $\tau=b(\varphi)$ we get
$$
\frac{\sigma(\rho) - A(\varphi)}{d(\rho)- \varphi}
= A'(\varphi) + a \big[\rho -b(\varphi) \big]
$$
and, by comparing (\ref{EL}) with (\ref{Deqbis}), we see that they are
indeed equivalent under the above condition.

As we emphasized, the rate function for SEP is obtained by
taking the supremum over $\varphi$ of $\mc G(\rho,\varphi)$, while for the
the KMP process we need to take the infimum. We can now realize that
this depends on the sign of $a$. Indeed for $a>0$ (as
in the KMP process) the functional $\mc G(\rho,\varphi)$ is concave in
$\nabla\varphi$ while it is convex for $a<0$ (as in the SEP).

\medskip
It is quite tempting to extend the previous derivation to a broader
class of models, possibly by a different definition of the trial functional
$\mc G$, however our attempts in this direction were not successful.

\subsection*{Acknowledgments}
It is a pleasure to thank C.\ Bernardin, A.\ De Sole,
G.\ Jona--Lasinio, C.\ Landim, and E.\ Presutti for useful discussions.
L.B. and D.G. acknowledge the support  COFIN MIUR 2002027798 and 2003018342.
The work of J.L.L. was supported by NSF Grant DMR 01-279-26 and AFOSR Grant
AF 49620-01-1-0154.

\end{document}